\documentclass[12pt,preprint]{aastex}
\begin{document}
\title{A New {\it FUSE} Survey of Interstellar HD}

\author{Theodore P. Snow\altaffilmark{1}, Teresa L. Ross\altaffilmark{1}, Joshua D. Destree\altaffilmark{1}, Meredith M. Drosback\altaffilmark{1}, Adam G. Jensen\altaffilmark{2}, Brian L. Rachford\altaffilmark{3}, Paule Sonnentrucker\altaffilmark{4}, and Roger Ferlet\altaffilmark{5}}

\altaffiltext{1}{Center for Astrophysics and Space Astronomy, Department of Astrophysical and Planetary Sciences, University of Colorado at Boulder, Campus Box 389, Boulder, CO 80309-0391, USA; tsnow@casa.colorado.edu; teresa.ross@colorado.edu; destree@casa.colorado.edu; meredith.drosback@colorado.edu} 
\altaffiltext{2}{NASA Postdoctoral Program Fellow, Goddard Space Flight Center, Code 665, NASA/GSFC, Greenbelt, MD 20771, USA; Adam.G.Jensen@nasa.gov} 
\altaffiltext{3}{Department of Physics, Embry-Riddle Aeronautical University, 3700 Willow Creek Road, Prescott, AZ 86301, USA; rachf7ac@erau.edu} 
\altaffiltext{4}{Department of Physics and Astronomy, Johns Hopkins University, 3400 North Charles Street, Baltimore, MD 21218, USA; sonnentr@pha.jhu.edu} 
\altaffiltext{5}{Institut d'Astrophysique de Paris, UMR7095 CNRS, Universite Pierre \& Marie Curie, 98bis Boulevard Arago, 75014 Paris, France; ferlet@iap.fr}

\begin{abstract}
We have used archival {\it FUSE} data to complete a survey of interstellar
HD in 41 lines of sight with a wide range of extinctions.  This follow up
to an earlier survey was made to further assess the utility of HD as a
cosmological probe; to analyze the HD formation process; and to see what
trends with other interstellar properties were present in the data.  We
employed the curve-of-growth method, supported by line profile fitting, to
derive accurate column densities of HD. We find that the N(HD)/2N(H$_2$)
ratio is substantially lower than the atomic D/H ratio and conclude that
the molecular ratio has no bearing on cosmology, because local processes
are responsible for the formation of HD.  Based on correlations with
E(B-V), H$_2$, CO, and iron depletion, we find that HD is formed in the
densest portion of the clouds;  the slope of the logN(HD)/log(H$_2$)
correlation is greater than 1.0, caused by the destruction rate of HD
declining more slowly than that of H$_2$; and, as a sidelight, that the
depletions are density dependent.  

\end{abstract}


\section{Introduction}

The H$_2$ isotopologue HD was first detected in {\it Copernicus} spectra by
Spitzer et al. (1973, 1974) and Morton (1975), and has subsequently been
observed in many sightlines by {\it FUSE}. The HD lines arising from the
lowest-lying rotational levels ($J$=0 and $J$=1) are far weaker than their
counterparts for H$_2$, but are in many cases still strong enough to be
saturated, requiring a curve-of-growth analysis for column density
determinations. To date HD analyses from {\it FUSE} spectra have been
published only for a few stars \cite{Ferlet,Lacour} though many more
detections reside in the {\it FUSE} archives. In addition, Lacour et
al. have re-analyzed {\it Copernicus} spectra for several stars, resulting
in a uniform survey of HD abundances in some 17 sightlines. The {\it
Copernicus} stars generally have A$_V$ values of 1.0 or less, while our
{\it FUSE} targets cover a range of A$_V$ from 0.1 to about 3 magnitudes.

The ratio of HD to H$_2$ --- more specifically, N(HD)/2N(H$_2$) --- in the
Lacour et al. survey ranges from a few times 10$^{-7}$ to several times
10$^{-6}$. This is somewhat higher than the values found earlier from {\it
Copernicus} data, but those values were based on two strong lines that were
almost always saturated.

We follow the convention of Lacour et al. (2005), and assert that, in a
region where all hydrogen (including deuterium) is in molecular form, the
atomic D/H ratio is the same as the N(HD)/2N(H$_2$).  The reasoning is that
the molecular fraction (N(HD)/N(H$_{total}$) = N(HD)/[N(H I) + 2N(H$_2$)])
reduces to N(HD)/2N(H$_2$) when H I can be neglected.  For every HD
molecule, there is one D atom, so the atomic ratio D/H is equal to
N(HD)/2N(H$_2$).  Note that this assumes that not only H I, but also atomic D,
can be neglected.  This simplifying assumption is justified in cloud cores,
but it may break down in regions where hydrogen (and deuterium) is only
partially in molecular form, which may include some of the less-reddened
lines of sight in our survey.  But even there, as shown in the next few
paragraphs, we can not use HD as an indicator of the D/H ratio anyway. 

It would be useful if the N(HD)/2N(H$_2$) ratio were a reliable indicator
of the atomic ratio D/H, because the latter is an important probe of the
early expansion rate of the universe and therefore valuable in cosmology.
An excellent summary of the observed atomic D/H ratio and its cosmological
implications can be found in Linsky et al. (2006). But for three
distinct reasons the comparison of HD to H$_2$ is difficult to apply to the
cosmological problem. 

First, H$_2$ begins to be self-shielding much before HD, as a function of
depth in a cloud. For example, H$_2$ becomes self-shielding at about
N(H$_2$) = 10$^{19}$  cm$^{-2}$ (which corresponds to N(HD) = 10$^{12-13}$
cm$^{-2}$) and a similar column density would be required for HD -- which is
incredibly unlikely. For HD to become self-shielding, the column density of
H$_2$ would have to be of order 10$^{25}$ cm$^{-2}$! Even infrared
observations of the quadrupole transitions at 2.4 $\mu$m would never reach
this column density. The fact that HD is not protected from radiative
dissociation acts to lower the N(HD)/2N(H$_2$) with respect to the atomic D/H
ratio in the denser portions of the lines of sight.   

Second, if formation of HD occurs on grain surfaces (in parallel with H$_2$ 
formation), the lower mobility of D  atoms on grains as compared to H
atoms, greatly reduces the formation rate of HD as compared to H$_2$,
acting to reduce N(HD)/2N(H$_2$) ratio relative to the D/H ratio.  If grain
formation of HD dominates, this would act to decrease the N(HD)/2N(H$_2$)
ratio over the atomic ratio. For a more complex and complete analysis of HD
formation on grain surfaces, see Cazaux et al. (2008).  

But third, HD has a gas-phase formation channel, through the "chemical
fractionation" reaction H$_2$ + D$^{+} \to$ HD + H$^{+}$ \cite{Watson,LePetit},
which tends to enhance the N(HD)/2N(H$_2$) ratio and depends on the
cosmic-ray ionization rate \cite{Black}.  The comparison of the
N(HD)/2N(H$_2$) ratio to the atomic D/H ratio can tell us which processes
are most important. 

In any case, the N(HD)/2N(H$_2$) ratio should not be expected to reflect
the D/H ratio (Lacour et al. 2005).  In general the N(HD)/2N(H$_2$) ratio
is not useful for determining the cosmological D/H ratio, although it may
be useful in constraining the rate of cosmic-ray ionization (which produces
D$^+$).  

This paper, then, presents a new and more complete survey of HD than was
available previously, to test formation theories and possible correlations
with other interstellar quantities. We have organized the paper into
sections as follows: Observation and Data Reduction, Data Analysis,
Determination of HD Column Densities, Correlations, Summary and Conclusions. 

\section{Observation and Data Reduction}

The {\it FUSE} mission covered a
wavelength range (905-1188 \AA) rich in ground state transitions, making it
one of the most useful tools to study the interstellar medium.  The 41
sightlines for this survey are archival {\it FUSE} spectra,
chosen to span different cloud types and physical conditions. The
sightline parameters used to select targets included color excess (0.11
$\le$ $E(B-V)$ $\le$ 0.83), the indicator of grain size R$_v$ (2.25 $\le$
$R_v$ $\le$ 4.76), molecular fraction (0.02 $\le$ f$_{H2}$ $\le$ 0.76),
and H$_2$ column density (19.07 $\le$ log N(H$_2$) $\le$ 21.11).  Table
\ref{table1} lists all of the relevant physical properties for the
sightlines included in this study.  Another selection criterion was a high
signal-to-noise ratio around the 1105.83 \AA\ HD line. This line is 100
times weaker than the rest of the HD absorption features measured, all of
which have similar oscillator strengths (HD f-values and wavelength data
from Abgrall \& Roueff 2006).  In many cases this weak line's
equivalent width is the only one falling on the linear portion of the curve
of growth, thus providing unambiguous information about the column density
as Lacour et al. found in their study.  

Because HD is known to be correlated with H$_2$ abundances \cite{Lacour},
our target list includes sightlines with H$_2$ column densities in the
literature.  H$_2$ column densities are taken from Shull et al. (in
preparation, 2008) and Rachford et al. (2002; in preparation, 2008). For
this study the {\it FUSE } data (Table \ref{table2}) were pre-calibrated
with version 3.1.4 or newer of the CALFUSE pipeline. For each observation
we used a cross correlation analysis to align individual exposures with
strong absorption features before co-adding them.  
 
In most cases we co-added the detector segments to increase the signal to
noise ratio before measuring the HD equivalent widths.  When combining the
detector segments one significant systematic flaw was found in the
data. The raw data from the LiF2A detector segment revealed the same
systematic detector defect due to a Type I dead zone (as described in
section 9.1.6 of The {\it FUSE} Instrument and Data Handbook) in all
sightlines in the vicinity of the weak 1105.83 \AA\ HD line; thus we
excluded that segment and only used LiF1B.

\section{Data Analysis}

To determine the column density of HD we measured equivalent widths for all 
possible HD lines (see Table \ref{eq_width}) and performed a curve of
growth (COG) analysis. All of the HD lines measured are transitions from 
the ground rotational state ($J$=0). While there are over 25 HD
absorption lines in the {\it FUSE} range, only seven are isolated enough
from other features to determine accurate equivalent widths.  To measure
the equivalent width we first defined a continuum on both sides of the
feature and normalized it with low order (1 to 3) Legendre polynomials. 
When there was no other absorption feature in the immediate vicinity we
integrated over the HD line while summing the flux errors provided by {\it
FUSE} in quadrature.  If there was another absorption feature blended with
the HD line, we fit the HD and interfering line with Gaussians (see next
section for details).  The errors in the width and depth of the Gaussian
were propagated into the equivalent width error.  The statistical error on
the equivalent width was summed in quadrature with the continuum placement
error. The continuum error was up to an order of magnitude smaller than the
statistical error in all cases. 

Of the seven lines, six have similar oscillator strengths (see Table
\ref{absorption_lines}); the seventh and weakest line is located at 1105.86
\AA\ and is very important for constraining the COG fit. The weak HD
line is difficult to measure in most cases and in other cases it is
impossible to measure at all. Without a line to measure we used the 
signal-to-noise ratio to set a two sigma upper limit on the equivalent
width using the equation: 
\begin{equation}
\sigma = \frac { \Delta \lambda  M^{1/2}} {(S/N)}
\end{equation}
where {$\Delta \lambda$} is the pixel scale, $S/N$ is the signal-to-noise
ratio, $M$ is the width of the feature (here {\it FUSE} resolution
elements ($\sim$9 pixels) are used since the width of the line will be
smeared out by the resolution), and $\sigma$ gives the one sigma error of the
equivalent width \cite{Jenkins}.

\subsection{Modeling the C I* Line}
The first obstacle in measuring the weak HD line is a very crowded patch of
continuum.  In a study of the sightline to $\zeta$ Oph, Morton (1978) found 
an unidentified absorption feature at 1105.92 \AA.  In the current study 
we have been unable to confirm or deny the existence of a weak feature near 
this wavelength; thus, if such a feature does exist it could be a source of 
systematic error.  Morton also rejected a feature at 1105.82 as being the
weak line of HD due to its large equivalent width. The conclusion of Lacour
et al. (2005) and the present study is that the feature at 1105.83 is
indeed HD and its strength is consistent with all of the other measured HD
features (see Figures \ref{plottwo}, \ref{plotthree}, \ref{plotfive} \&
\ref{plotsix} containing HD COG's). Shortward of the HD line, by 0.13 \AA,
is an absorption feature from the first excited state of carbon as
discussed by Lacour et al. In sightlines where the C I* and HD lines were
blended we modeled the C I* and divided it from the spectrum before
measuring the HD line. 

Removing this blended C I* line turned out to be more challenging than
originally thought.  The f-value of the 1105.73 C I* line,
previously measured as 0.0113 by Morton (1978), did not fit the curve of
growth set by the other C I* lines.  This is probably due to uncertainty in
the adopted f-value of this weak line.  The previous value was based on a
single measurement of the line in the spectrum of $\zeta$ Oph.  So to
accurately model the C I* line we used existing archival data to make an
empirical estimate of the f-value, based on more sightlines.

\begin{figure}
\begin{center}
\scalebox{.5} [.5]{\includegraphics{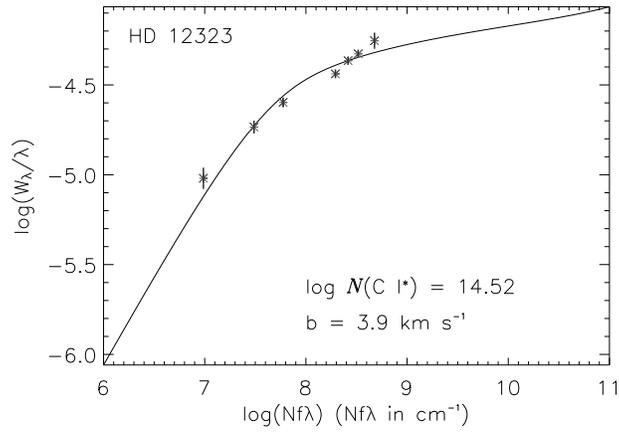}}
\figcaption{Carbon I* Curve of growth.  \label{plotone}}
\end{center}
\end{figure}

\begin{figure}
\scalebox{.75} [1]{\includegraphics[angle=90]{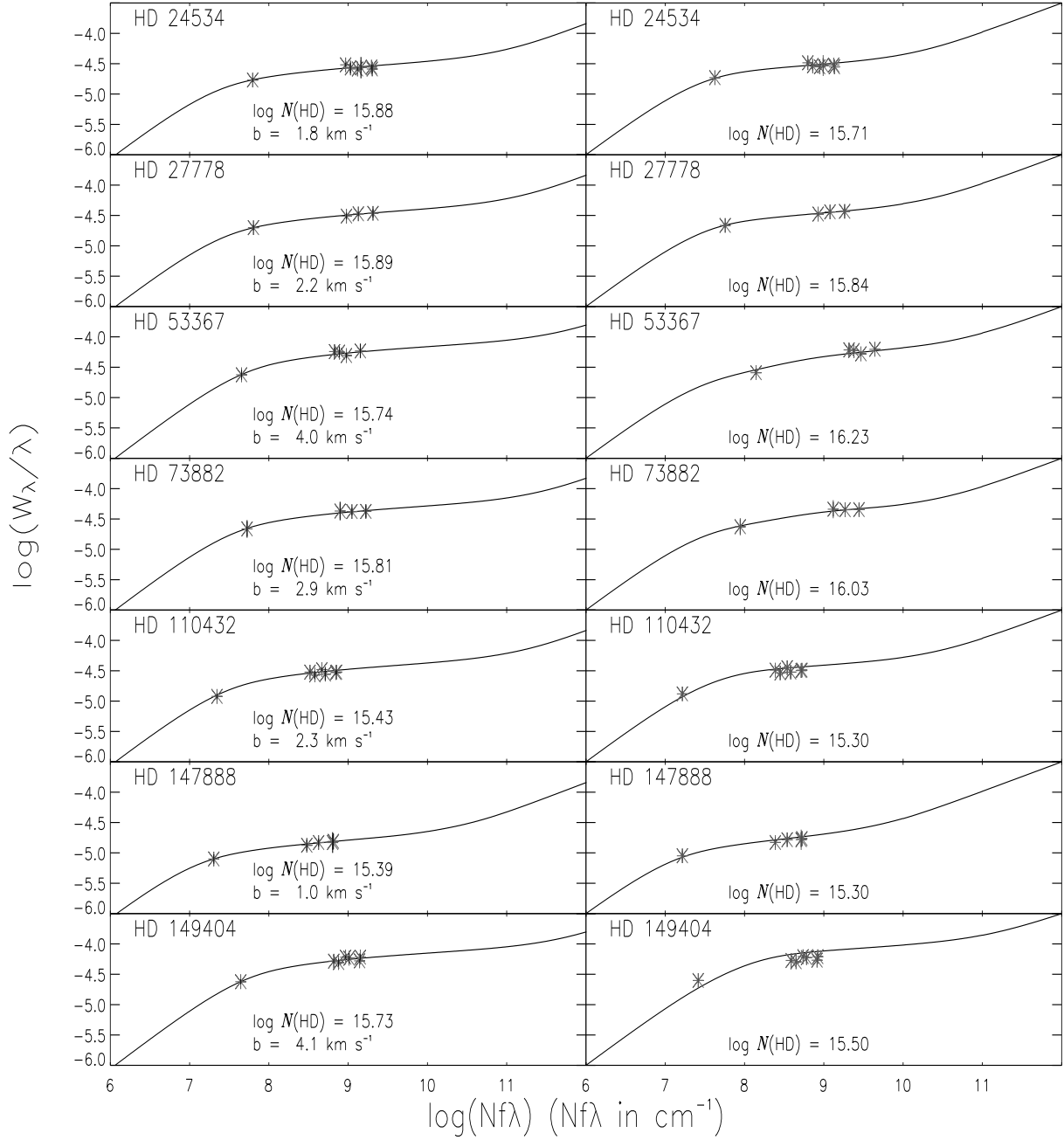}}
\caption{Single Component (left) and Multiple Component (right) Curves of
Growth for HD 24534 through HD 149404. \label{plottwo}}
\end{figure}

\begin{figure}
\scalebox{.75} [1]{\includegraphics[angle=90]{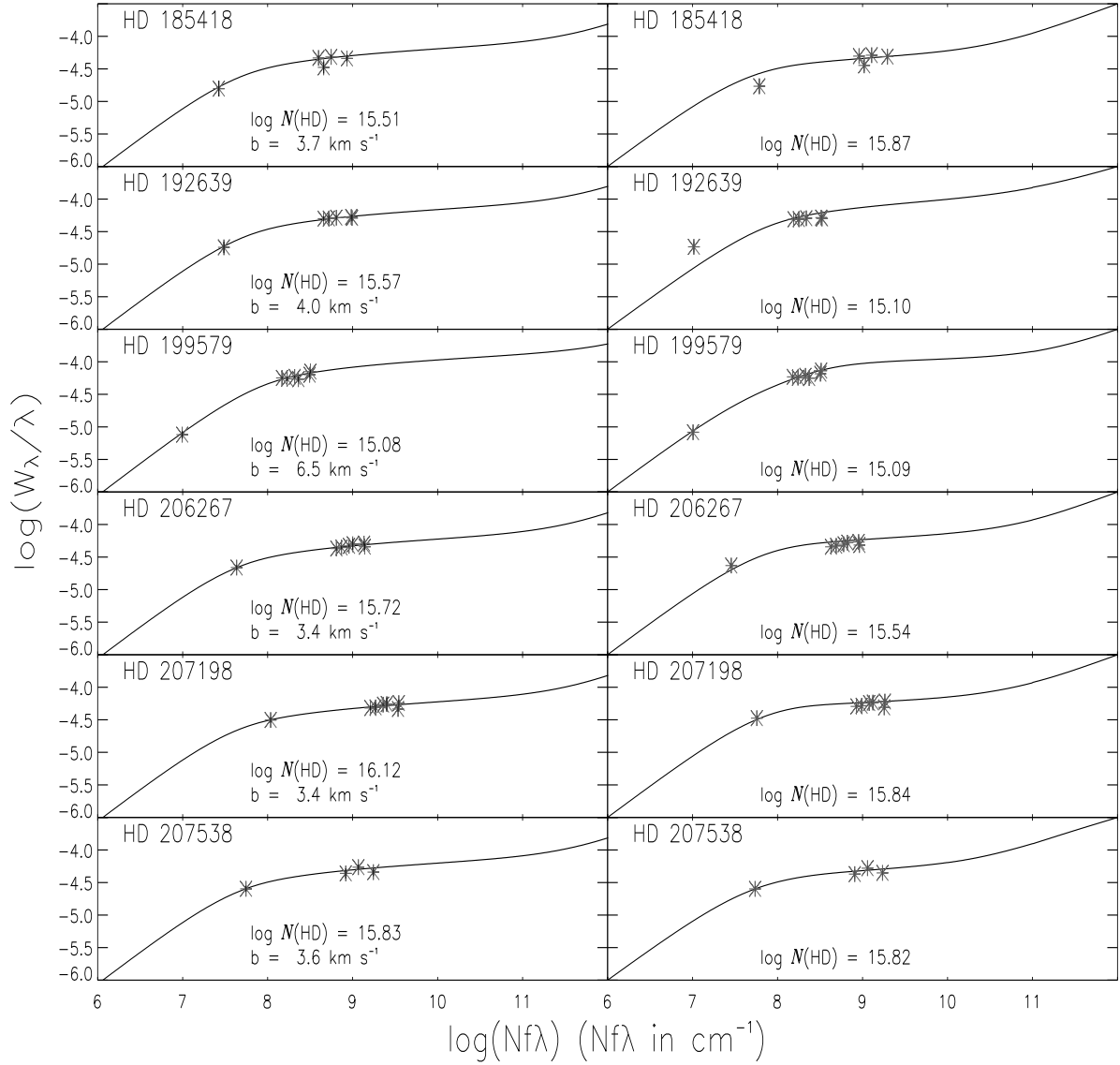}}
\caption{Single Component (left) and Multiple Component (right) Curves of
Growth for HD 185418 through HD 207538. \label{plotthree}}
\end{figure}

To accomplish this task we found {\it Copernicus}, {\it FUSE}, and/or
STIS data in sightlines where the 1105 \AA\ C I* line was well 
defined, not
heavily blended with other lines in the region (HD 12323, $o$ Per, HD
207538, HD 210839) and measured the equivalent widths of all well defined C
I* lines, excluding the 1105 \AA\ line.  For each sightline we then fit
these equivalent widths to a single component curve of growth varying both
$b$-value and column density (see Figure \ref{plotone} for an example).
Once a best fit curve of growth was found, the f-value of the 1105 \AA\
line was calculated using its measured equivalent width.  Uncertainty in
the f-value was assigned based on errors in the measured equivalent width
of the 1105 \AA\ line. Calculated f-values and uncertainties for each
sightline can be found in Table \ref{tab_CIfvalue}.  The results from all
sightlines were combined using a weighted average and the final f-value was
found to be 0.0062$^{+0.0015}_{-0.0010}$.

The C I* lines used in modeling are in Table \ref{absorption_lines}.  We
used STIS data when available, along with {\it FUSE} data to measure
the C I*  absorption features.  Using the best fit column density and
$b$-value of C I*, we generated a Voigt profile of the 1105.73 \AA\ C I*
line.  This profile was then convolved with the Gaussian instrumental
profile (assuming a resolution of 20 km s$^{-1}$) and divided out of the
spectrum, allowing us to measure the HD 1105.83 \AA\ line.

We began the current study using C I* line f-values from Morton's
compilation (1991).  However, in light of newer theoretical f-values from
Zatsarinny \& Froese Fischer (2002) and Froese Fischer (2006), it was
important to determine whether differences due to revised f-values would
substantially affect our HD column densities. Repeating the calculation of
the C I* 1105.73 \AA\ line f-value yielded only a very small change (about 
three percent) from the previously calculated f-value. All C I* column
densities were also recalculated using the newer f-values.  Total column
densities did not vary greatly from those using the Morton (1991) f-values.
In all but three cases differences in C I* column density using the two
sets of f-values differed by 0.1 dex or less.  The three most discrepant
cases were HD 73882, HD 101436 and HD 149404.  For these three cases the
equivalent width of the weak HD line was remeasured and found to be well
within the 1$\sigma$ errors of the previous measurements.  Similarly, in
all three cases rerunning the COG fit yielded only small changes in the HD
column density that were within the 1$\sigma$ error bars of the previous
measurements (in all cases the difference was less than 0.1 dex). Thus, we
did not think it necessary to repeat all of the C I* modeling, and
remeasure the weak HD line, as other errors far exceed any error introduced
by minor changes in some of the C I* f-values, and changes in C I* column 
densities were not substantial in the vast majority of cases. All reported C
I* column densities are those using the newer f-values from Zatsarinny \&
Froese Fischer (2002) and Froese Fischer (2006).

\section{Determination of HD Column Densities}
Once the influence of C I* had been removed, we were able to do a
curve-of-growth analysis to determine the column density,  but we used
profile fitting in a few cases as a check.  We did two different COG
analyses;  multiple component COG's for a sub-set of 13, and single
component COG's for all 41 sightlines.  Both are described below in
subsection 4.1 'Curve of Growth Analysis'. The profile fitting is 
described in subsection 4.2.

\subsection{Curve of Growth Analysis}
In many cases multiple cloud components exist along a single line of sight
increasing the difficulty in accurately measuring the column density. 
Having knowledge of such structure is most important when the 
equivalent widths fall on the flat portion of the curve. We can create a
multiple component COG, but first we need to find a suitable tracer of HD. 

HD has been found to be correlated to H$_2$ \cite{Lacour}. Similarly CH is
traced by H$_2$ because of the chemistry: the creation of CH in the diffuse
ISM is directed through a series of gas phase reactions with H$_2$ 
(Danks, Federman \& Lambert 1984; van Dishoeck \& Black 1989; Magnani et 
al. 1998). Since CH and HD are both correlated to H$_2$ abundances, there
is an inferred correlation between CH and HD, which has been verified by
Lacour et al.  
 
\begin{figure}
\begin{center}
\scalebox{.4} [.36]{\includegraphics[angle=90]{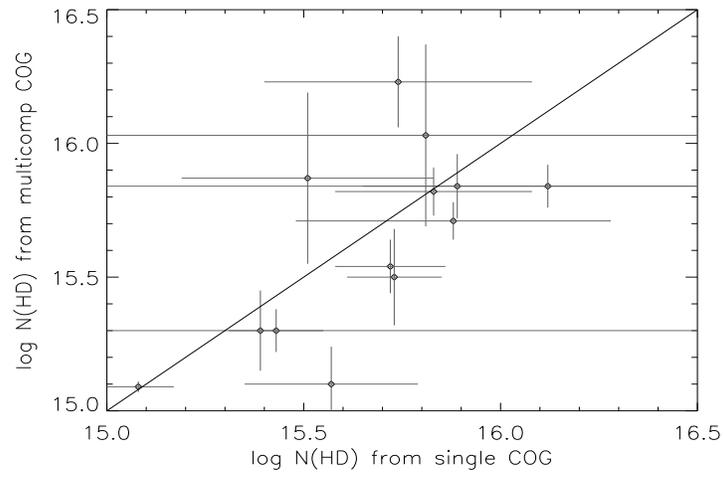}}
\caption{Column Density solutions from a multiple component curve of growth 
compared to a single component curve of growth. \label{plotfour}}
\end{center}
\end{figure}

\begin{figure}
\scalebox{.8} [1]{\includegraphics[angle=90]{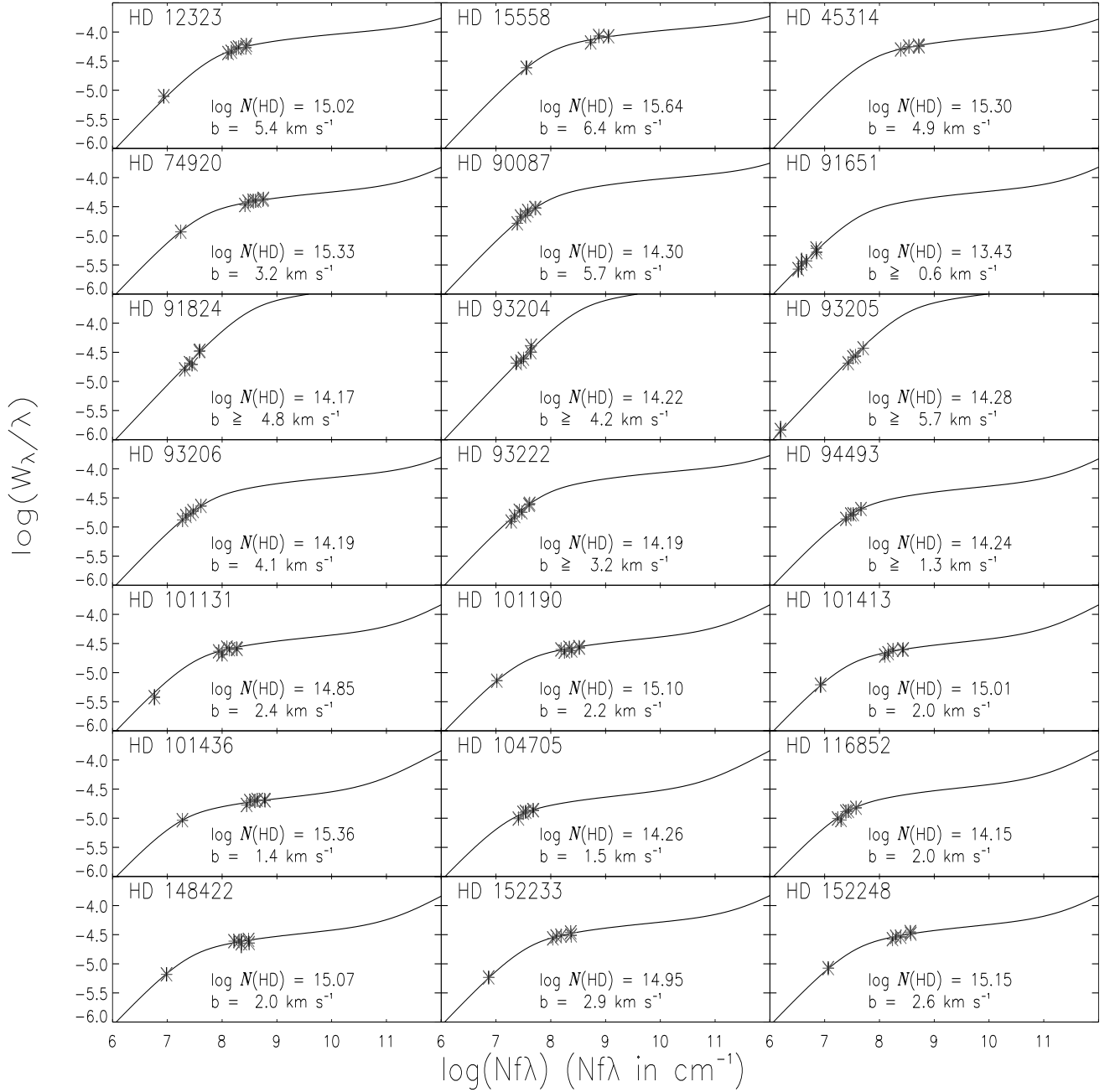}}
\caption{Single Component Curves of Growth with column densities and $b$-values 
for HD 12323 through HD 152248. \label{plotfive}}
\end{figure}

\begin{figure}

\scalebox{1} [.9]{\includegraphics{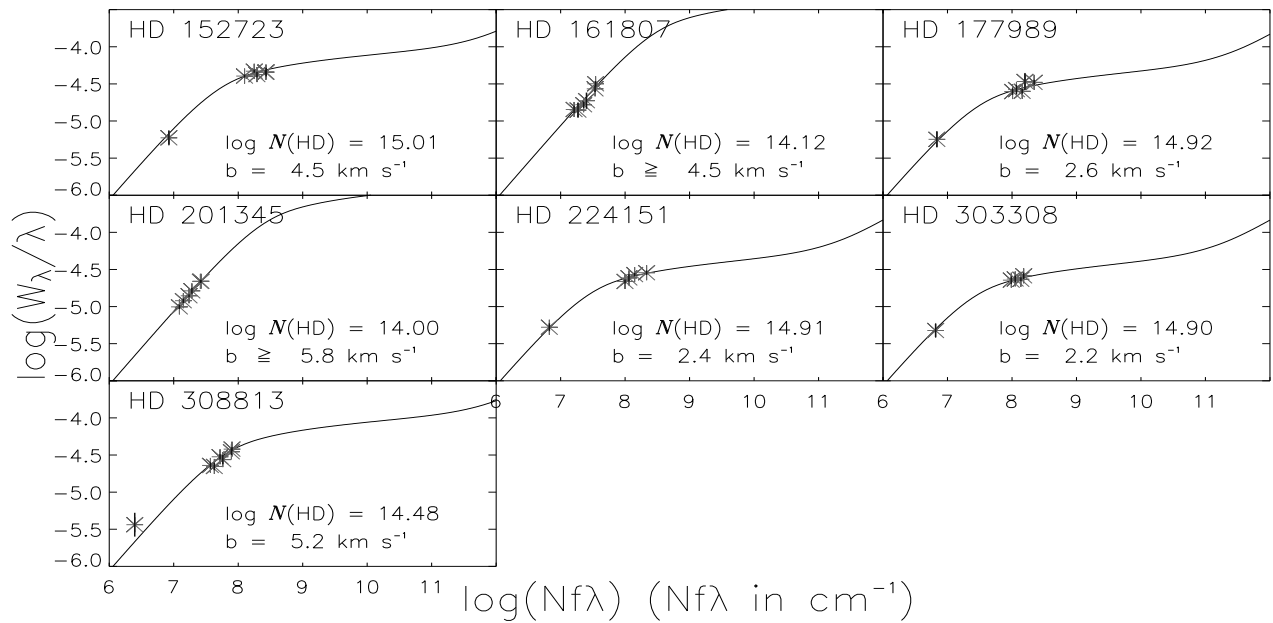}}
\caption{Single Component Curves of Growth with column densities and $b$-
values for HD 152723 through HD 308813. \label{plotsix}}
\end{figure}

For thirteen sightlines we had high resolution CH data (Welty, private
communication) which we used to define a velocity structure for HD. From
the measured CH line-of-sight velocity structure, we generated multiple
component COG's.  The HD equivalent widths were fit to the curve varying
only the total column density.  For comparison the data were also fit to a
single component COG that found a best fit of both $b$-value and column
density by minimizing the ${\chi}^2$.  Figures \ref{plottwo} and
\ref{plotthree} show the side by side comparison of the COG analyses, and
Figure \ref{plotfour} compares the column densities and errors from the two
methods.   

The COG's from the two methods fit the data in all but three cases. The
side by side comparison show a clear mis-fit of the multiple component COG
for HD 149404, HD 185418, HD 192639, even though the column densities are
within one-sigma errors for the first two targets.  The discrepancy arose
because a multiple component COG assumes the best fit velocity structure of
CH is the actual structure of the HD and does not allow for errors. Varying
the $b$-values on the multiple component COG within their one sigma errors
gave solutions that were consistent with the data and the single component COG.

With the reasonable assurance that our single component COG model is good
we accounted for any other systematic errors that we could not quantify by
setting the reduced $\chi ^2$ of the best-fit curve equal to one.  In 22
sightlines the reduced $\chi ^2$ was already less than one, thus the errors
were unchanged; for the remaining 19 cases the assumed good fit scaled up
the errors in equivalent width and column density.

The consistency between the results of the single and multiple component
COG demonstrates that a single component model is a good approximation when
we do not have component information.  Thus, we are justified in combining
the HD results for the 13 targets with component information and the 28
targets without; see Figures \ref{plotfive} and \ref{plotsix} for single
component COG's and Table \ref{colden} for column densities. 

Solving for column density in the COG analysis was more complicated when we
only had an upper limit on the 1105 \AA\ line.  We used the limit on the
weak line to constrain the column density by throwing out solutions beyond
the upper limit.  The limit was not used in calculating the $\chi ^2$ of
the fit.  

Lower limits on the $b$-value were determined when all the points fell on
the linear portion of the COG. The $b$-value controls where the curve will
transition from the linear to the saturated portion. Since the transition
cannot occur lower than the data points a two sigma lower limit is set (HD
91824, HD 93204, HD 93205, HD 94493, HD 161807, and HD 201345.)

\subsection{Profile Fitting}

Profile fitting of the HD lines was done for five lines of sight to verify
the COG column density measurements.  All the available HD
lines were fit simultaneously so that a single best fit column density and
$b$-value would be measured.  First the continuum surrounding each HD feature
was normalized by fitting a low order Legendre polynomial to the continuum
on both sides.  Shifts in the central velocity of the HD features were
noticed from one line to another, (with each detector segment showing
different velocity shifts, usually on the order of $\sim$10 km s$^{-1}$) so
the lines were each fit with a single Gaussian and the central wavelength
corrected so that each feature would be at its rest wavelength.  Profiles
were modeled assuming a single Voigt profile with column density $N$, line
width $b$ and central velocity $v$. These profiles were then convolved with
a Gaussian instrumental profile. We used the non-linear least-squares curve
fitting algorithm MPFIT by C. Markwardt\footnote{\tt
http://cow.physics.wisc.edu/$\sim$craigm/idl/idl.html} to find the best fit
parameter values.  MPFIT is a set of routines that uses the
Levenberg-Marquardt technique to minimize the square of deviations between 
data and a user-defined model.  These routines are based upon the MINPACK-1
Fortran package by More' and collaborators\footnote{\tt
http://www.netlib.org/minpack}. 

Attempts were also made to fit multiple components when previous
measurements of CH existed.  However, even when the velocity structure was
fixed by the CH structure, individual component column densities did not
converge to reasonable answers.  Usually one component of the fit dominated
while the other component(s) were an order of magnitude or two lower in
column density (differing greatly from the relative strengths of the CH
components).  Because of this, we report only the single component fits in
Table \ref{profile_vsCOG}. 

\begin{figure}
\begin{center}
\scalebox{.4} [.36]{\includegraphics{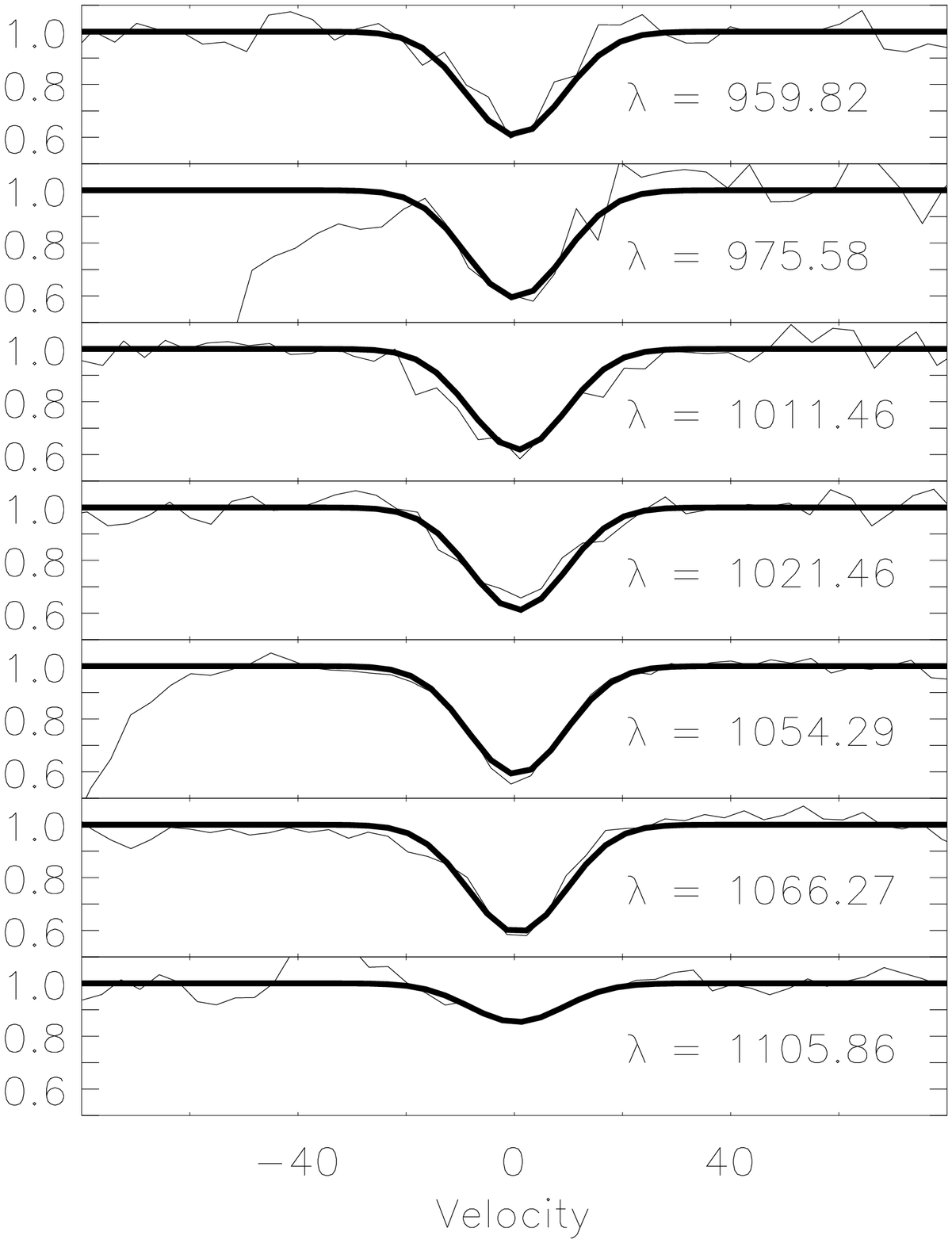}}
\caption{Profile fit of HD 110432's seven absorption lines, simultaneously
solving for column density (log N(HD) = 15.32) and $b$-value  (2.32).  The
velocity is set to the rest wavelength due to varying velocity solutions
from different detector segments. \label{plotseven}} 
\end{center}
\end{figure}

For all five sightlines we find that the column densities and $b$-values
for both the curve of growth and profile-fitting method are within standard
errors of one another (including HD 192639, our previously discrepant
case).  We also find that the best fit profiles match the absorption
profiles in the data well (see Figure \ref{plotseven}).  Thus, we are
confident that our column densities measured through curve of growth
fitting are reliable.

\subsection{Comparisons with previous results}

We are able to compare our column densities with the previous HD
survey of Lacour et al. (2005). This survey had 7 stars observed
with {\it FUSE} and 10 more lines of sight observed years before, with 
{\it Copernicus}.  The comparison of our results are in Table
\ref{French_compare}. 

We re-analyzed the 7 {\it FUSE} targets that Lacour measured to check
whether our method was consistent with their results.  Our method differs
in a few small ways.   We re-derived the C I* f-value (as described above),
so our de-blending of the weak HD line is different.  Also we weighted each
line equally in the COG fit, while the Lacour group doubly-weighted the
optically thin HD line.  Even with these difference our column densities
are consistent within the uncertainties.

We also did single-component curves of growth for every sightline,
which the Lacour group did not, and we found consistent column densities 
with the multicomponent results.  This allowed us to expand our survey with 
confidence, in the end obtaining accurate column densities for 41 stars.

Previous HD measurements made by Spitzer et al. (1974) are included in Figure
\ref{ploteight} and the correlation analysis.  The plot includes the upper
limits for $\delta$ Per and $\lambda$ Ori and the measurements for $\pi$
Sco, $\delta$ Sco, $\sigma$ Sco, and $\gamma$ Ara.

\section{Correlations}
One way to derive information about the formation mechanism of HD, and to
test theoretical chemical models, is to see how various interstellar
quantities correlate with HD column densities. Before we do that, a few
words of caution are in order. Of necessity, we can measure only integrated
line-of-sight quantities, not localized to the same portion of the lines of
sight. For example, any correlations of molecular quantities versus dust
indicators or H I are probably not informative, because both dust and H I
are everywhere, inhabiting every region from diffuse to dark clouds. We
assume, with good reason, that the observable HD and H$_2$ (along with CH) are
confined to diffuse molecular and translucent clouds (as defined by Snow
\& McCall 2006), so we should not expect to find good correlations with either
dust or H I. If we do find good or decent correlations, that would indicate
that diffuse molecular and translucent clouds dominated the lines of sight
in our survey. On the other hand, those quantities that do correlate well
with each other would show that they arise in the same portion of the line
of sight.  All correlations can be found in Table \ref{correlate},
including some not illustrated.  

\begin{figure}
\scalebox{.4} [.4]{\includegraphics[angle=90]{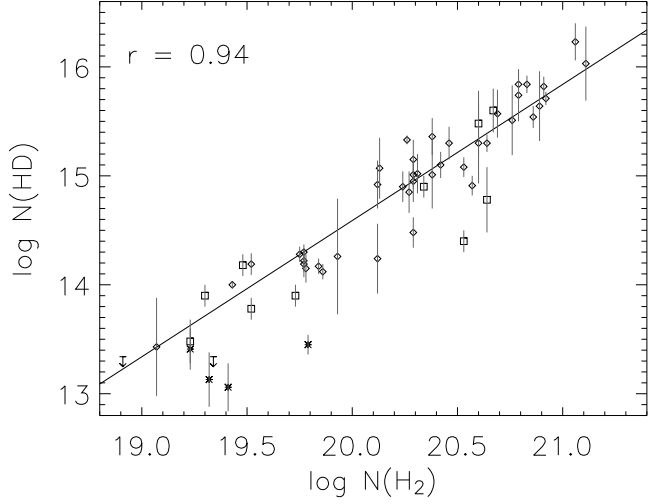}}
\caption{{Correlation plots of log N(HD) versus log N(H$_2$). The line is 
a weighted least-squares fit. Square plot symbols represent {\it
Copernicus} data analyzed by Lacour et al. 2005, diamonds represent this
study, asterisk and upper limit represent {\it Copernicus} data reported by
Spitzer et al. 1974.} \label{ploteight}} 
\end{figure}

The first correlation of interest is the one between HD and H$_2$, which is 
shown in Figure \ref{ploteight}. There we see a very good correlation,
suggesting that the formation of HD depends on the local abundance of H$_2$, 
or that both are formed by the same or similar processes.  Because H$_2$
shoots up in column density due to the self-shielding about when N(H$_2$)
reaches 10$^{19}$ cm$^{-2}$, we checked for a change in slope around that
point. We see more scatter below log N(H$_2$) = 19.5 and no significant
change in slope (below 19.5, the slope is 1.55, based on a fairly narrow
range; above, it is 1.33).  Considering all of our data points together
gives an overall slope of 1.25 and a correlation coefficient is 0.94. 

The significant departure from a slope of 1.00 in this correlation can be
interpreted in two different ways: either HD is being formed at a 
faster rate than the formation rate of H$_2$ in the portion of the clouds
where both are forming; or HD is destroyed at a lesser rate than the 
destruction rate of H$_2$ in those regions.  This latter explanation may be
preferred, because of optical depth effects and how they affect the
photodestruction rates.  For most of the observed range in A$_V$, the H$_2$
$J$ = 0 and 1 lines are damped, while the corresponding lines of HD lines are
becoming saturated - but not damped.  In this range in A$_V$ (between A$_V$
about 0.1 and 0.8), the rate of HD destruction is rapidly declining, while the
photodestruction rate of H$_2$ is only slightly declining \cite{LePetit}. 
This situation mimics the appearance of relative rate of formation favoring
HD, and that explains the slope of the N(HD)/N(H$_2$).

The alternative explanation of the slope in Figure \ref{ploteight}, that HD
is being formed faster than H$_2$, could only be viable if some additional
formation process were at work, in addition to the standard gas-phase
reaction network.  And that would imply HD formation on grain surfaces
competes with the gas-phase mechanism. But the detailed analysis of HD
formation on grains by Cazaux et al. (2008) explores conditions in which
grain formation of HD could be important, and our observed lines of sight
do not meet the criteria outlined by Cazaux et al. So our explanation, that
HD has started to be destroyed at a lesser rate, is more likely. 

This finding of a greater slope than 1.00 seems to be inconsistent with the
gas-phase  models of Le Petit et al. (2002), who predict that the ratio of
HD over H$_2$ should decline over our entire range in molecular fraction.
Their Figure 4 shows the ratio N(HD)/2N(H$_2$) as a function of the molecular
fraction of hydrogen, and this figure shows a steady decline with
increasing molecular fraction. Only above a molecular fraction of 0.9
does the ratio start to increase.  The study by Le Petit et al. only considers
a single cloud, whereas real sightlines cover many clouds in most cases.
But it is difficult to see how this helps reconcile the model with the
observations, because the situation only gets worse when multiple clouds,
which collectively form the total column density, are considered. The model
would be correct only if all the molecular regions along the lines of
sight, each having a molecular fraction $\ge$ 0.9, contained {\it all} the
HD and H$_2$, with no spillover to less dense regions.  

\begin{figure}
\scalebox{.32} [.32]{\includegraphics[angle=90]{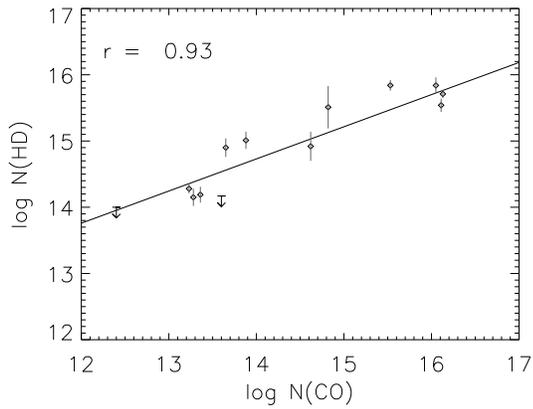}}
\caption{{Correlation of log N(CO) with log N(HD). CO data from Burgh et
al. 2007}\label{plotnine}} 
\end{figure}

Figure \ref{plotnine} shows the correlation between the column densities of
HD and CO (the CO data were taken from Burgh et al. 2007). Here we see a very
good correlation (coefficient 0.93), which we take as an indication that
related processes account for both HD and CO.  Since CO is formed by
gas-phase reactions (e.g Kaczmarczyk 2000a; 2000b; and references cited 
therein), we again conclude that HD is also likely formed by gas-phase
reactions. The ion-neutral reaction H$_2$ + D$^+ \to$ HD + H$^+$
\cite{LePetit}, with the ion D$^+$ being created by cosmic-ray impacts, is
commonly thought to be the primary gas-phase formation channel for HD. 

\begin{figure}
\scalebox{.32}[.32]{\includegraphics[angle=90]{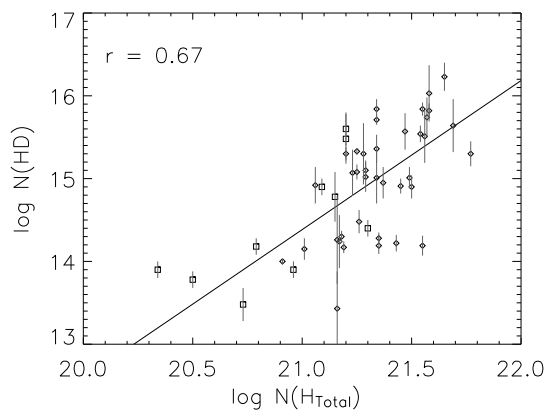}}
\caption{{Correlation plots of log N(HD) versus log N(total Hydrogen).  Square
plot symbols represent {\it Copernicus} data analyzed by Lacour et
al. 2005, diamonds represent this study.} \label{plotten}}
\end{figure}

\begin{figure}
\scalebox{.32} [.32]{\includegraphics[angle=90]{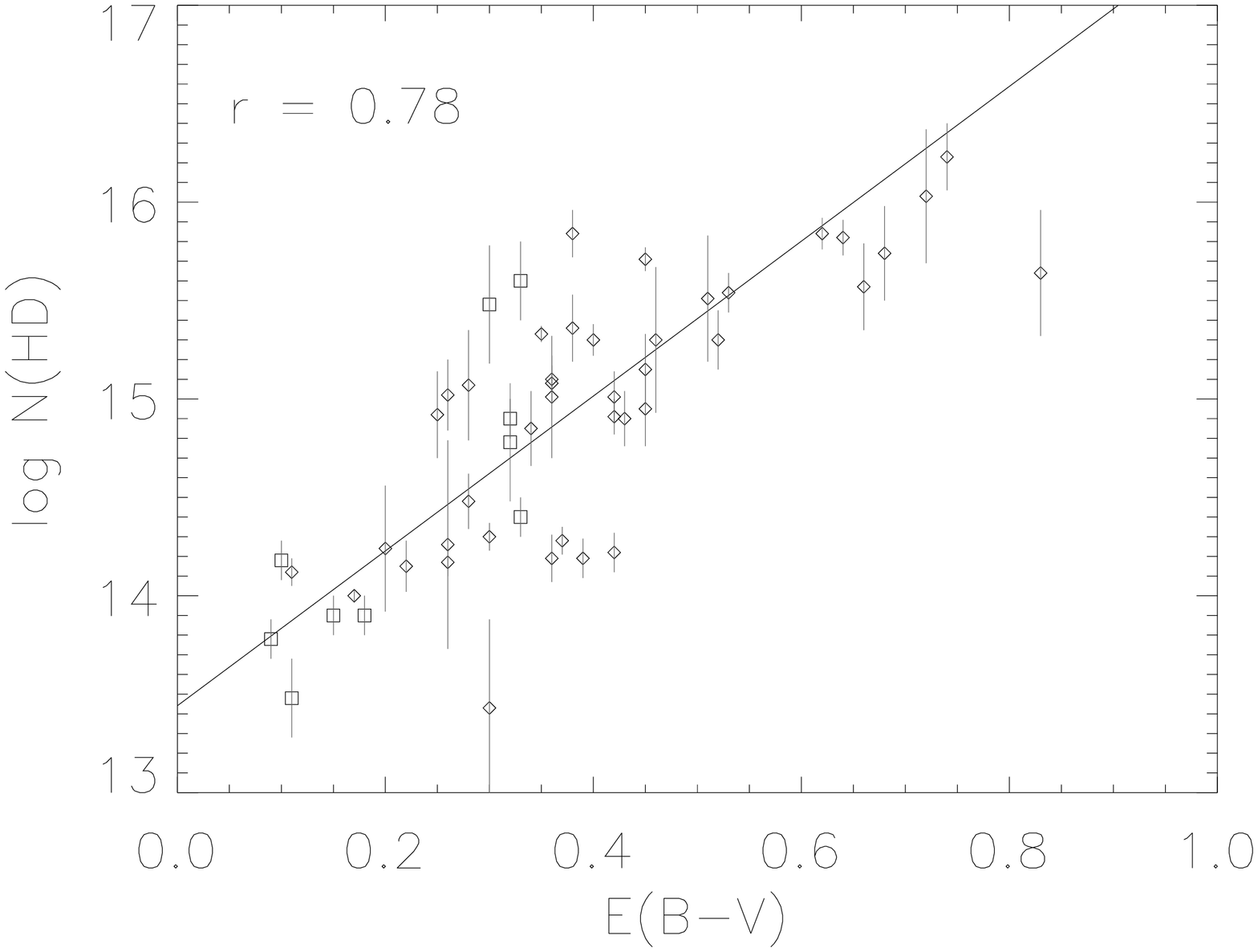} \quad \includegraphics[angle=90]{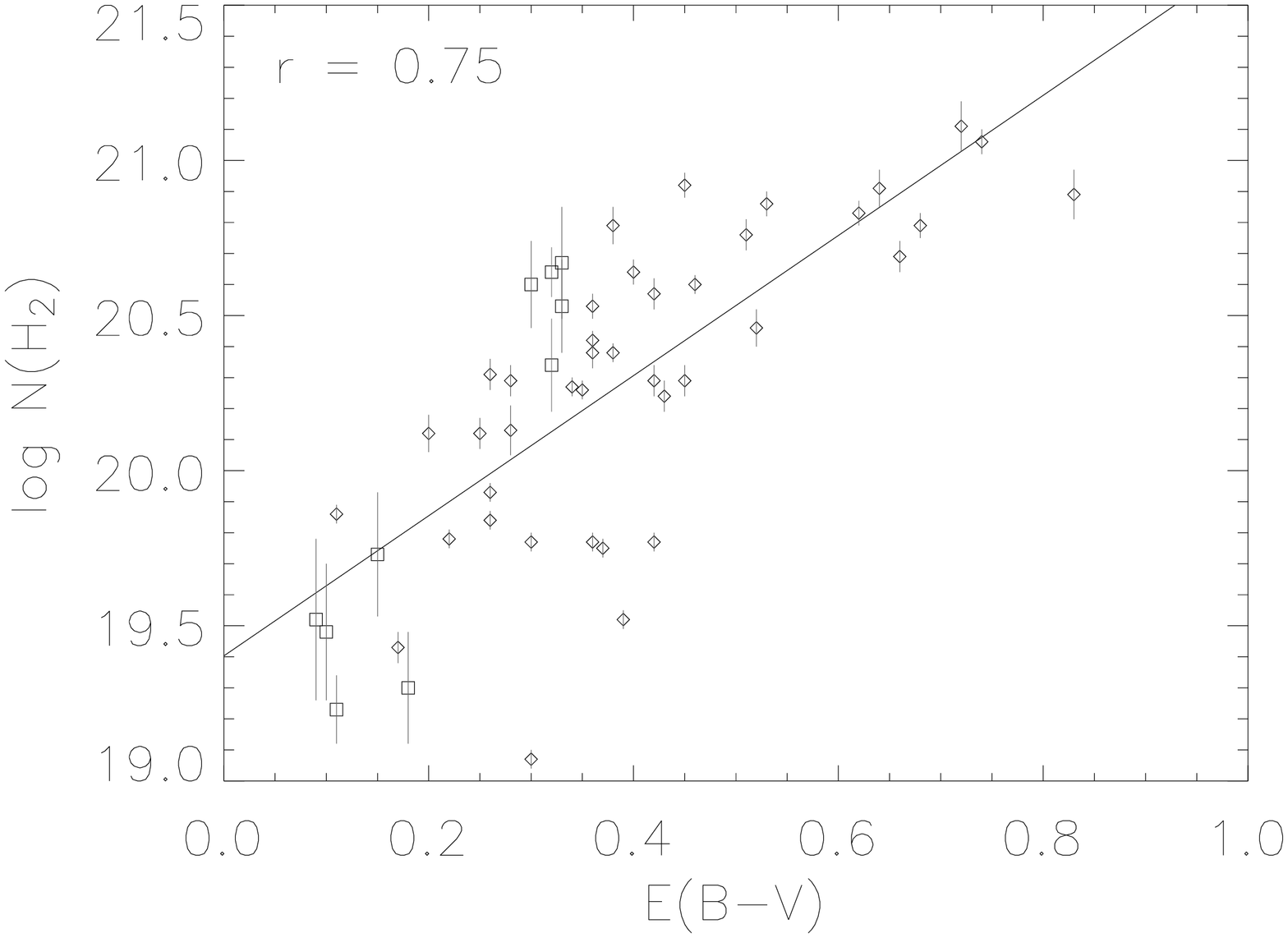}}
\caption{{log N(HD) and log N(H$_2$) versus reddening,
E(B-V). Symbols same as Figure \ref{plotten}.} \label{ploteleven}}
\end{figure}

\begin{figure}
\scalebox{.32} [.32]{\includegraphics[angle=90]{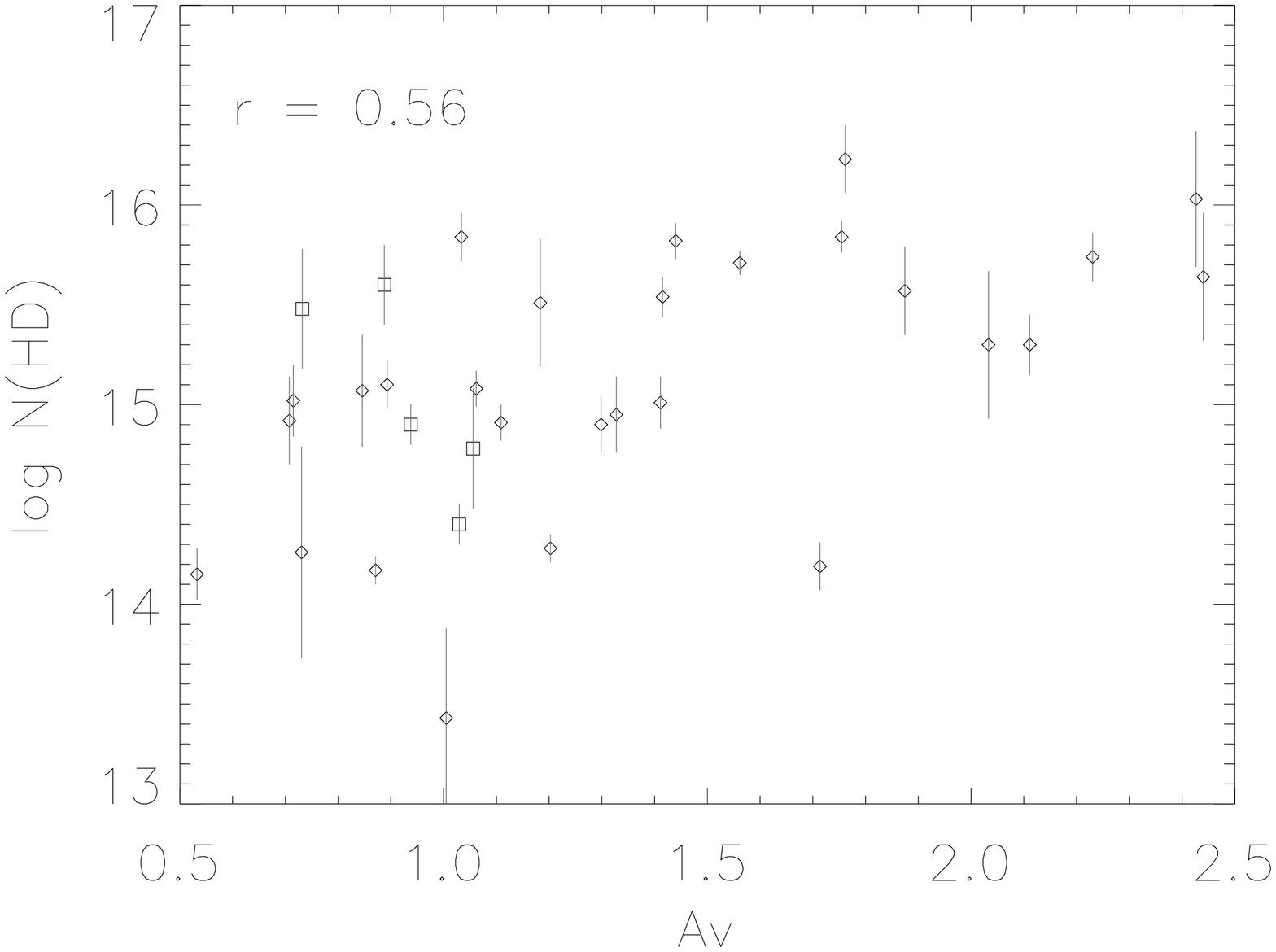} \quad \includegraphics[angle=90]{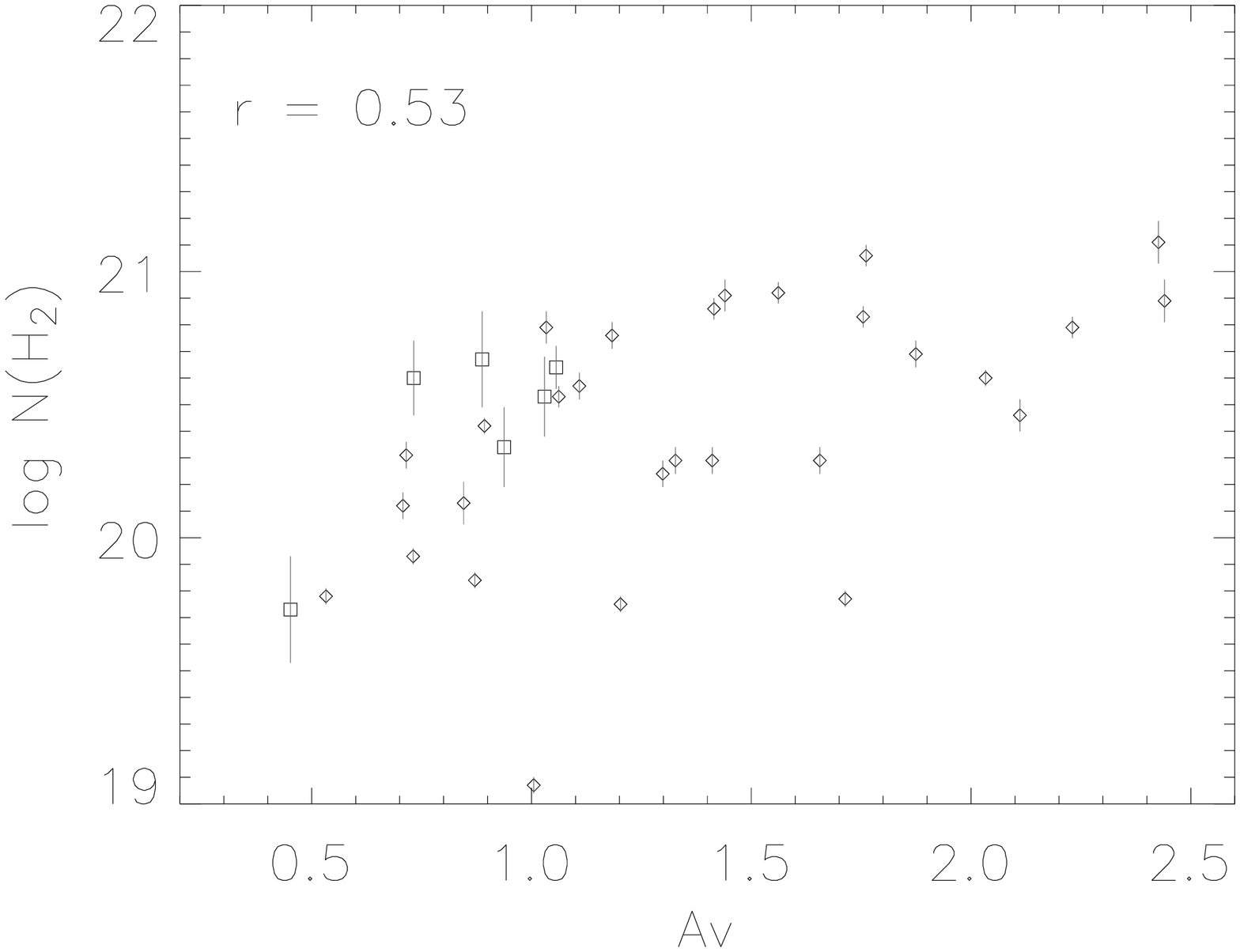}}
\caption{{log N(HD) and log N(H$_2$), versus extinction (A$_V$).  Symbols same as Figure \ref{plotten}.}
\label{plottwelve}} 
\end{figure}

To test our assertion that HD and H$_2$ do not inhabit the whole line of sight
in most cases, we show the correlations of both with quantities that are
expected to arise over the entire sightline. In Figure \ref{plotten} we show
the correlations of HD with the dust parameters total H (i.e., N(H I) +
2N(H$_2$)); Figure \ref{ploteleven} shows the correlations of H$_2$ and HD
with E(B-V); and Figure \ref{plottwelve} shows the correlations of H$_2$
and HD with  A$_V$. These correlations do not differ significantly from
each other, except there is a hint that both HD and H$_2$ go with E(B-V)
slightly better than they correlate with either total H, or A$_V$.  Neither
HD or H$_2$ correlate significantly with R$_V$.
 
The correlations of HD and H$_2$ with E(B-V) are better than those with H
I, telling us something surprising and perhaps useful: that the grains
responsible for differential extinction tend to be in the densest parts of the
observed lines of sight, where HD and H$_2$ are present.  The grains
responsible for E(B-V) are thought to be comparable in size to the 
wavelengths being scattered, so we conclude that other grains in the lines
of sight may have a different size distribution.

We tested correlations of both H$_2$ and HD against the molecular fraction,
given by f = N(H$_2$)/N(H$_{Total}$), and found a fairly good correlation
(see Table \ref{correlate}). Neither should be a surprise, because they
just show that H$_2$ and HD exist primarily in the densest portions of the
lines of sight, as expected.   

\begin{figure}
\begin{center}
\scalebox{.32} [.32]{\includegraphics[angle=90]{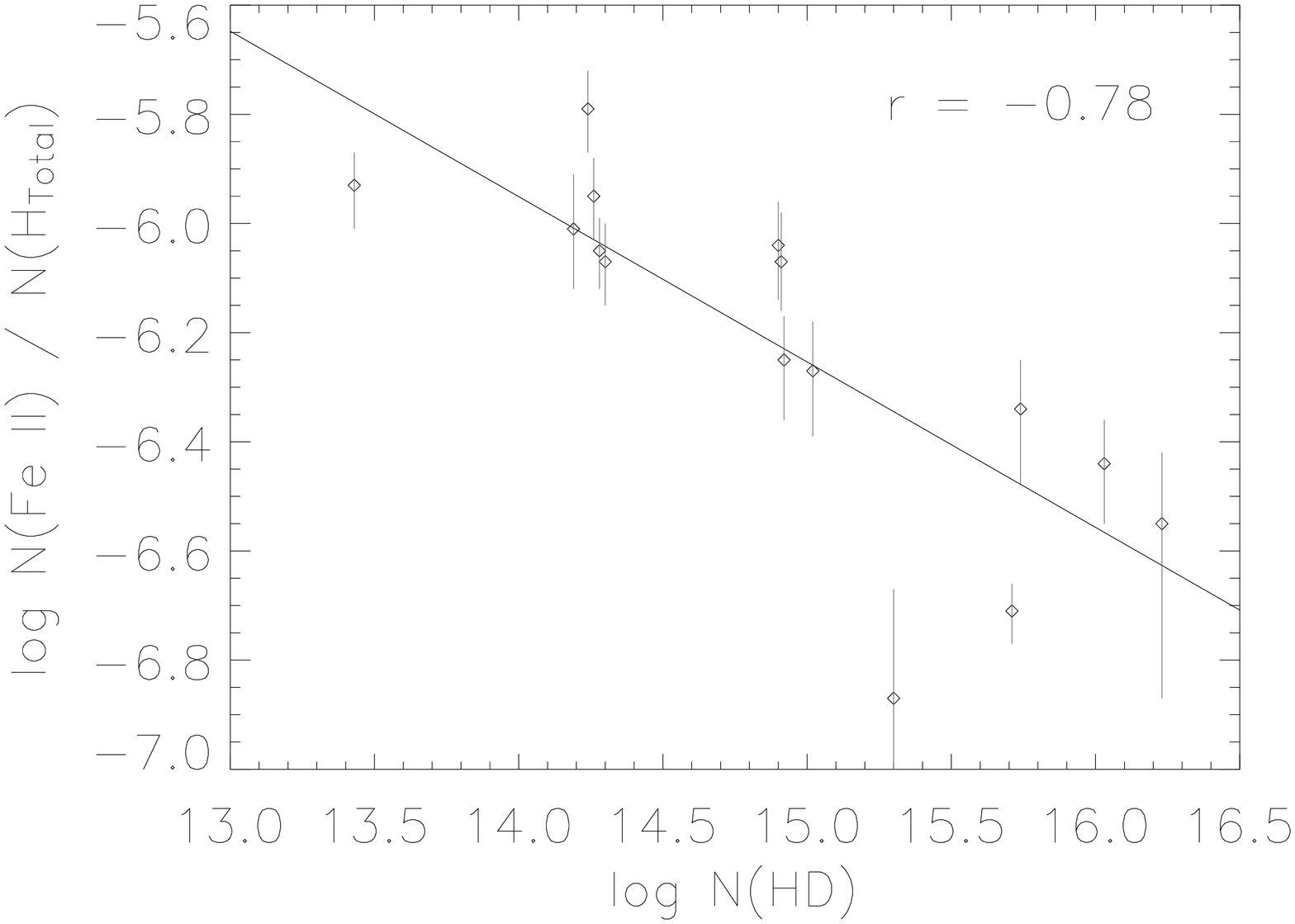} \quad \includegraphics[angle=90]{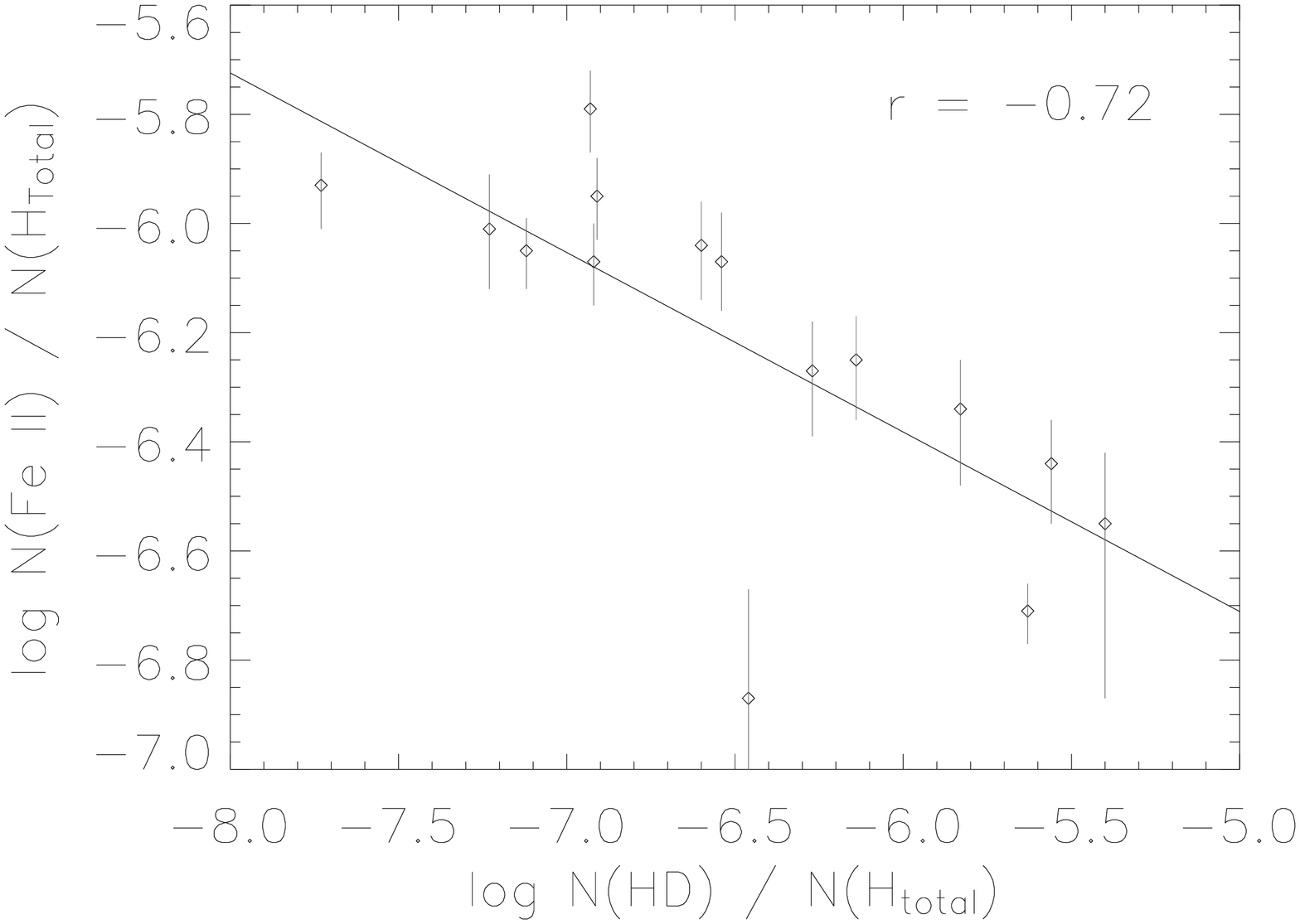}}
\vspace{.5 cm}
\scalebox{.32} [.32]{\includegraphics[angle=90]{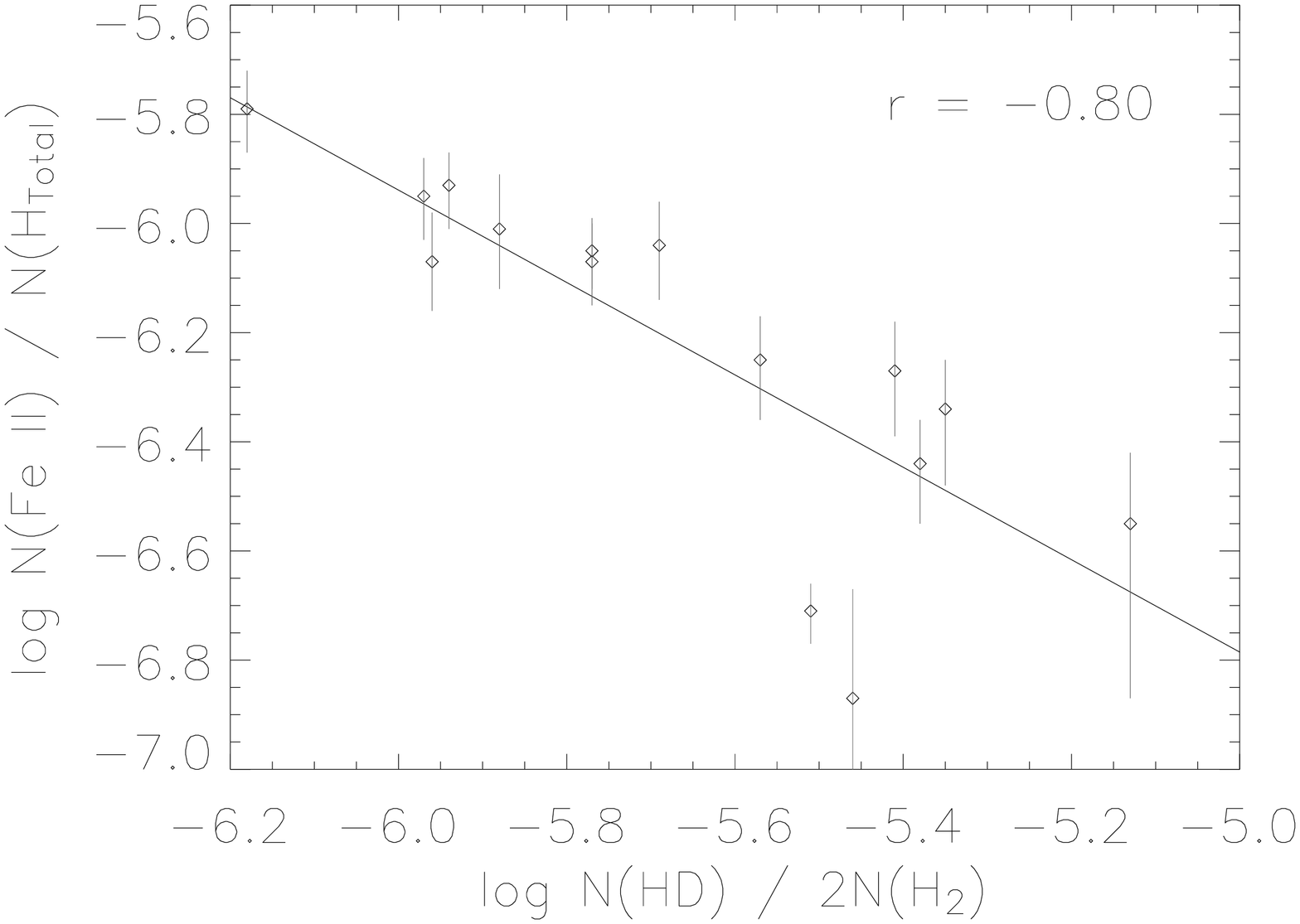}}
\caption{{Correlations with iron depletions, a density indicator.  Iron
data in Table \ref{table1} originally reported by Jensen \& Snow 2007.} 
\label{plotthirteen}}
\end{center}
\end{figure}

Finally, we have looked at correlations between iron depletion, as shown by
the ratio of iron to total hydrogen, and we see several interesting
things. These correlations are shown in Figure \ref{plotthirteen}. We could
have considered correlations with depletions of other elements, but we have
more complete data on iron for our lines of sight, and we expect other
refractory elements to follow more or less the same trends as iron. The
iron column densities come from Jensen \& Snow (2007), and are listed in
Table \ref{table1}. 

The three correlations we considered are consistent with the idea that the
depletions are density dependent. Remembering that an increasingly negative
value of the Fe/H ratio means more depletion of iron, we see that the
depletion is growing with indicators of density such as the HD column
density, the fractional ratio of HD (i.e., N(HD)/N(H$_{Total}$); and the
ratio of HD to H$_2$ (see Figure \ref{plotthirteen}). The indicators we see
here,  showing that depletion increases with density, is consistent with
the finding of Burgh et al. (2007), who found that the depletions correlate
with CO, another density indicator.

\section{Summary and Conclusions}

We have derived column densities of interstellar HD for 41 lines of sight,
covering a wide range of total extinctions, total hydrogen column density,
and differential extinction E(B-V). We went to considerable effort to
reduce error in the derived column densities, which were ultimately found 
using the curve-of-growth method, after measuring the equivalent
widths. One challenge we faced, and successfully resolved, was that the
weakest line of HD, detectable only toward highly-reddened stars, was
contaminated by a weak C I* line. We had to model and remove the C I* 
line in those cases. Where possible, we checked our results against other
published column densities, with generally good agreement. 

We found values of the logarithmic ratio of N(HD)/2N(H$_2$) ranging from
-6.18 to -5.13 -- generally much lower than the galactic atomic D/H ratio
and even lower with respect to the extragalactic (or the primeval) ratios
\cite{Linsky}.  This says two things: (1) the enhanced photodissociation of HD
over H$_2$ dominates and lowers the N(HD)/2N(H$_2$) ratio; and (2) our initial
assumption that we can not use N(HD)/2N(H$_2$) ratio as a cosmological test
is confirmed, and only the atomic ratio gives a true value of the
primordial production of deuterium (if that; see Linsky et al. 2006).

In order to see whether our results are consistent with extant chemical
models, we examined correlations of HD with other interstellar parameters,
finding the following: (1) HD is declining more slowly than the destruction 
rate of H$_2$; (2) N(HD)
correlates well with N(CO); and (3) the depletion of iron (and probably
other refractory elements) is enhanced in the densest portions of the lines
of sight.    The model of Le Petit et al. (2005) of the formation of HD is
inconsisten with the observations, unless individual dense clouds along the 
lines of sight contain {\it all} the HD and H$_2$.

\acknowledgements

We are very grateful to the anonymous referee, who stimulated us to 
revise this paper in many useful ways. We are grateful to several
colleagues (including especially Dan Welty and 
Mike Shull) who gave us their advice, not to mention their data; and to
NASA, which funded this research through grants NNG05GA85G and NAS5-98043.
Also thanks to Ms. Lynsi Aldridge who helped with the early analysis of the
data.

\clearpage

\begin{deluxetable}{lccccccclcc}
\tablecolumns{11}
\tablewidth{0pc}
\tabletypesize{\scriptsize}
\tablecaption{Sightline Properties \label{table1}}
\tablehead{Star  &E(B-V) &$R_V$ &Ref.\tablenotemark{a} &Molec. &log
  N(H$_2$) &log N(CH) & log N(H$_{Tot}$) &Ref.\tablenotemark{a} &log N(Fe II)\tablenotemark{b} &log N(C I*)\tablenotemark{c}  \\
(HD) &&&&Fract. &(cm$^{-2}$) &(cm$^{-2}$) &(cm$^{-2}$) &&(cm$^{-2}$) &(cm$^{-2}$) }

\startdata
 
12323	&0.26	&2.75	&1 	&0.21	&20.31 	& ...	&21.29	&4  &15.02 &...	  \\
15558	&0.83	&2.94	&2	&0.32	&20.89	& ...	&21.69	&4  &...   &...	  \\
24534   &0.45	&3.47	&3	&0.76	&20.92	&13.57 	&21.34	&5  &14.63 &14.98 $\pm$ 0.02 \\
27778  	&0.38	&2.72	&3	&0.56	&20.79	&13.48 	&21.34	&5  &...   &15.15 $\pm$ 0.32 \\
45314	&0.46	&4.42	&1	&0.42 	&20.60 	& ...	&21.28	&4  &...   &...	   \\
53367	&0.74	&2.38	&3	&0.52	&21.06	& ...	&21.65	&3 &15.08 &15.16 $\pm$ 1.83 \\
73882  	&0.72  	&3.37 	&3	&0.67 	&21.11	&13.57	&21.58	&5  &15.15 &16.35 $\pm$ 0.64 \\
74920	&0.35	&...	&...	&0.21	&20.26	&...	&21.25	&4  &...   &...   \\
90087  	&0.30	&...	&...	&0.08	&19.77	& ...	&21.18	&4  &15.15 &...    \\ 
91651	&0.30	&3.35	&1	&0.02 	&19.07	& ...	&21.16	&4  &15.23 &...   \\
91824 	&0.26	&3.35	&1	&0.09	&19.84	& ...	&21.19	&4  &...   &...   \\ 
93204	&0.42	&...	&...	&0.04	&19.77	& ...	&21.43	&4  &...   &...   \\
93205	&0.37	&3.25	&1	&0.05	&19.75	& ...	&21.35	&4  &15.35 &14.13 $\pm$ 0.05  \\
93206  	&0.39	&...	&...	&0.03	&19.52	& ...	&21.35	&4  &...   &14.13 $\pm$ 0.17  \\ 
93222	&0.36	&4.76	&1	&0.03 	&19.77	& ...	&21.55	&4  &...   &13.77 $\pm$ 0.04  \\ 
94493	&0.20	&...	&...	&0.18	&20.12	& ...	&21.17	&4  &15.38 &...    \\
101131	&0.34	&...	&...	&...	&20.27	& ...	&...	&4  &...   &...    \\
101190	&0.36	&2.48	&1	&0.27 	&20.42	& ...	&21.29	&4  &...   &14.54 $\pm$ 0.11  \\
101413  &0.36	&...	&...	&0.22	&20.38	& ...	&21.34	&4  &...   &...    \\
101436	&0.38	&...	&...	&0.22	&20.38	& ...	&21.34	&4  &...   &14.49 $\pm$ 0.22  \\
104705	&0.26	&2.81	&1	&0.12	&19.93	& ...	&21.16	&4  &15.22 &...   \\
110432 	&0.40	&...	&...	&0.55	&20.64	&13.19 	&21.20	&5  &...   &14.55 $\pm$ 0.26  \\
116852	&0.22	&2.42	&1	&0.12	&19.78	& ...	&21.01	&4  &...   &...    \\
147888	&0.52	&4.06 	&3	&0.10	&20.46	& ...	&21.71	&3 &14.89 &15.02 $\pm$ 0.10  \\
148422	&0.28	&3.02 	&1	&0.16	&20.13	& ...	&21.23	&4  &...   &...    \\
149404	&0.68	&3.28 	&3	&0.33	&20.79	& ...	&21.57	&3 &15.23 &14.86 $\pm$ 0.09  \\ 
152233	&0.45	&2.95	&1	&0.17	&20.29	& ...	&21.37	&4  &...   &...    \\
152248	&0.45	&3.68	&1	&...	&20.29	& ...	&...	&4  &...   &...   \\
152723	&0.42	&3.36 	&1	&0.13	&20.29	& ...	&21.49	&4  &...   &...    \\ 
161807	&0.11	&...	&...	&...	&19.86	& ...	&...	&4  &...   &...    \\
177989 	&0.25	&2.83 	&1	&0.23	&20.12	& ...	&21.06	&4  &14.81 &14.52 $\pm$ 0.03  \\
185418 	&0.51	&2.32	&3 	&0.47	&20.76	&13.12 	&21.56	&5  &...   &14.35 $\pm$ 0.08  \\ 
192639	&0.66	&2.84	&3 	&0.32 	&20.69	&13.45 	&21.47	&5  &...   &14.54 $\pm$ 0.08  \\ 
199579	&0.36	&2.95	&3	&0.38	&20.53	&13.36 	&21.25	&5  &...   &14.78 $\pm$ 0.12  \\ 
201345	&0.17	&...	&...	&0.03	&19.43	& ...	&20.91	&6  &...   &...    \\
206267	&0.53	&2.67	&3	&0.42	&20.86  &13.41 	&21.54	&5  &...   &14.91 $\pm$ 2.09  \\
207198  &0.62	&2.42	&3	&0.38	&20.83  &13.56 	&21.55	&5  &...   &15.18 $\pm$ 0.16  \\
207538	&0.64	&2.25	&3	&0.43	&20.91	&13.63 	&21.58	&5  &...   &15.23 $\pm$ 0.15  \\ 
224151	&0.42	&2.64	&2	&0.26	&20.57	& ...	&21.45	&4  &15.38 &14.34 $\pm$ 0.06  \\ 
303308	&0.43	&3.02	&1	&0.11	&20.24	&... 	&21.50	&4  &15.46 &14.38 $\pm$ 2.61  \\ 
308813	&0.28	&...	&...	&0.22	&20.29	& ...	&21.26	&4  &...   &...    \\

\enddata
\tablenotetext{a}{References: (1) Valencic et al. 2004;  (2) Rachford, private
communication; (3) Rachford in preparation, 2008; 
\\(4) Shull in preparation, 2008; (5) Rachford et al. 2002; (6) Burgh et
al. 2007}
\tablenotetext{b}{Jensen \& Snow 2007}
\tablenotetext{c}{This study}

\end{deluxetable}

\begin{deluxetable}{lllllll}
\tablecolumns{7}
\tablewidth{0pc}
\tabletypesize{\scriptsize}
\tablecaption{Observation data \label{table2}}
\tablehead{Star   &Stellar &V &Dist\tablenotemark{a} &Ref.	&{\it FUSE} Data ID   	&STIS Data ID \\
(HD) &Type &(mag)&(pc)& && }

\startdata
12323	&O9V	&8.92 	&3900	&1	& P1020202000  	&\\
15558	&O5IIIf	&7.95	&2187  	&2	& P1170101000  	&\\
24534	&O9.5pe	&6.10	&398   	&2	& P1930201000 	&o66p01020 \\
27778	&B3V	&6.33	&220   	&3	& P1160301000	&o59s01010 \\
45314	&O9pe	&6.60	&799    &2	& P1021301000 	&\\
53367  	&BoIV:e	&6.94	&247    &3	& P1161101000   &\\
73882  	&O9III	&7.24	&925    &1	& P1161302000   &\\
74920	&O8	&7.54	&1497   &2	& P1022601000   &\\
90087  	&O9.5III&7.80	&2716   &2	& P1022901000   &\\
91651	&O9Vn	&8.87	&3500   &1 	& P1023102000   &\\
91824  	&O7V	&8.15	&4000  	&3 	& A1180802000   &\\ 
93204  	&O5Vf	&8.48	&2630  	&2	& P1023501000   &\\
93205	&O3V	&7.76	&2600  	&1	& P1023601000 	&o4qx01030 \\
93206  	&O9.7Ibn&6.24	&2512  	&2	& P1023401000   &\\
93222  	&O7IIIf	&8.11	&2900  	&1	& P1023701000 	&o4qx02030 \\
94493	&B0.5Iab&7.27	&3327  	&2	& P1024101000   &\\
101131 	&O6Vf	&7.16	&709   	&3	& P1024901000   &\\ 
101190 	&O6Vf	&7.27	&2399  	&2	& P1025001000   &\\
101413 	&O8V    &8.33	&2399  	&2	& P1025301000   &\\
101436 	&O6.5V	&7.58	&2399  	&2	& P1025401000  	&o6lz51010 \\
104705 	&B0Ib	&7.83	&3898  	&2	& P1025701000   &\\
110432 	&B1IIIe	&5.24	&301   	&3	& P1161401000   &\\
116852	&O9III	&8.49 	&4760  	&2	& P1013801000   &\\
147888 	&B3	&6.78	&136   	&3	& P1161501000 	&o59s05010 \\
148422	&B1Ia	&8.64	&8836  	&2	& P1015001001   &\\
	&&&&				& P1015002001 	&\\
	&&&&				& P1015003001 	&\\
149404 	&O9Iae	&5.47	&1380  	&2	& P1161702000   &\\
152233	&O6III	&6.59	&1905  	&2	& P1026702000   &\\
152248	&O7Ibnfp&6.11	&1758  	&4	& P1026801000   &\\
152723 	&O6.5IIIf &7.31	&1905  	&2	& P1027102000   &\\
161807	&B0IIIn	&7.01	&383   	&3	& P1222302000   &\\
177989 	&B0III 	&9.34	&4909  	&2	& P1017101000   &\\
185418 	&B0.5V	&7.45	&950   	&5	& P1162301000 	&o5c01q010 \\
192639 	&O7Ib(f) &7.11	&1100  	&1	& P1162401000 	&o5c08t010 \\
199579 	&O6Ve 	&5.96	&794   	&2	& P1162501000   &\\
201345	&O9V	&7.75	&1907  	&2	& P1223001000   &\\ 
206267 	&O6.5V	&5.62	&850   	&1	& P1162701000   &\\
207198 &O9.5Ib-II &5.96 &832   	&2	& P1162801000 	&o59s06010 \\
207538 	&O9V 	&7.31	&832   	&2	& P1162902000 	&o63y01010 \\
224151 	&B0.5II &6.05	&1355  	&2	& P1224103000 	&o54308010 \\
303308 	&O3Vf	&8.21	&2630  	&2	& P1221602000 	&o4qx04010 \\
308813 	&O9.5V	&9.32	&2398  	&2	& P1221903000   &\\
\enddata
\tablenotetext{a}{Reference: (1) Savage et al. 1985; \  (2) Diplas \&
Savage 1994; \  (3) The Hipparcos and Tycho Catalogues;
  \ (4) Raboud et al. 1997;  \  (5) Sonnentrucker et al. 2003}

\end{deluxetable}

\clearpage

\begin{deluxetable}{lcc}
\tablecolumns{3}
\tablewidth{0pc}
\tabletypesize{\scriptsize}
\tablecaption{Previously Reported HD Column Densities Comparison \label{French_compare}}
\tablehead{ &Previous Values\tablenotemark{a} & Our Values \\
Star	&log N(HD)	&log N(HD) \\
(HD)	& (cm$^{-2}$)	&(cm$^{-2}$)}

\startdata
27778	&15.51 $^{+0.30}_{-0.33}$	&15.84 $\pm$ 0.12 \\
73882	&15.76 $^{+0.21}_{-0.38}$	&16.03 $\pm$ 0.34 \\
110432	&15.28 $^{+0.14}_{-0.17}$	&15.30 $\pm$ 0.08 \\
185418	&15.63 $^{+0.16}_{-0.13}$	&15.51 $\pm$ 0.32 \\
192639	&15.18 $^{+0.20}_{-0.27}$	&15.57 $\pm$ 0.22 \\
206267	&15.32 $^{+0.23}_{-0.28}$	&15.54 $\pm$ 0.10 \\
207538	&15.70 $^{+0.31}_{-0.28}$ 	&15.82 $\pm$ 0.09 \\
\enddata
\tablenotetext{a}{Values from Lacour et al. 2005}
\end{deluxetable}

\begin{deluxetable}{lccccccc}
\tablecolumns{8}
\tablewidth{0pc}
\tabletypesize{\scriptsize}
\tablecaption{Equivalent Widths of HD lines \label{eq_width}}
\tablehead{Star     &W$_{959.82}$ &W$_{975.58}$ &W$_{1011.46}$  &W$_{1021.46}$ &W$_{1054.29}$  &W$_{1066.27}$
&W$_{1105.86}$ \\
 (HD) &(m\AA)&(m\AA) &(m\AA) &(m\AA) &(m\AA) &(m\AA) &(m\AA)  }

\startdata

12323  &43.9 $\pm$ 3.0  &53.0 $\pm$ 3.8 &60.4 $\pm$ 3.3 &55.1 $\pm$ 2.5	&55.0 $\pm$ 2.3	&46.4 $\pm$ 3.0	& 8.7 $\pm$ 2.5 \\
15558  &...		&...		&...		&86.0 $\pm$ 19.0&90.1 $\pm$ 12.5&70.3 $\pm$ 10.2&26.9 $\pm$ 7.6 \\
24534  &25.8 $\pm$ 5.2  &26.5 $\pm$ 10.7&26.2 $\pm$ 4.2 &28.4 $\pm$ 2.8	&27.3 $\pm$ 1.8	&32.3 $\pm$ 2.3	&18.8 $\pm$ 1.4 \\
27778  &... 		& ...		&34.7 $\pm$ 5.6	&... 		&35.4 $\pm$ 2.7	&32.9 $\pm$ 4.4	&22.0 $\pm$ 2.2 \\
45314  &...		&...		&57.0 $\pm$ 1.7	&59.9 $\pm$ 1.4	&59.2 $\pm$ 1.2	&53.4 $\pm$ 1.1	& $<$ 25.7 \\
53367  &53.8 $\pm$ 10.5 & ...		&...		&59.8 $\pm$ 15.8&51.9 $\pm$ 5.7	&61.2 $\pm$ 8.6	&26.3 $\pm$ 4.9 \\
73882  &...		&...		&...		&43.0 $\pm$ 10.1&44.2 $\pm$ 6.3	&45.6 $\pm$ 16.0&24.0 $\pm$ 8.0 \\
74920  &37.0 $\pm$ 3.3	&38.9 $\pm$ 3.8	&43.7 $\pm$ 1.9	&42.2 $\pm$ 1.5	&42.4 $\pm$ 1.7	&36.0 $\pm$ 1.6	&13.0 $\pm$ 0.8 \\
90087  &20.6 $\pm$ 2.0 	&26.1 $\pm$ 3.6 &30.3 $\pm$ 1.4 &30.7 $\pm$ 1.5	&23.7 $\pm$ 1.2	&17.5 $\pm$ 1.1	& $<$ 4.9 \\
91651  & 3.2 $\pm$ 1.4	&...		&5.3 $\pm$ 0.9	& 6.2 $\pm$ 1.3	& 3.9 $\pm$ 0.8	& 2.9 $\pm$ 1.0	&... \\
91824  &15.4 $\pm$ 2.3	&19.1 $\pm$ 2.6 &34.1 $\pm$ 6.9	&33.7 $\pm$ 2.3	&21.7 $\pm$ 1.7	& ...		& $<$ 3.8 \\
93204  &19.8 $\pm$ 5.0 	&23.9 $\pm$ 4.1 &41.5 $\pm$ 4.4 &32.4 $\pm$ 6.8	&23.1 $\pm$ 2.2	&...		& $<$ 4.5 \\
93205  &19.8 $\pm$ 1.3	&26.6 $\pm$ 1.5	&37.6 $\pm$ 1.4	&...  	 	&27.4 $\pm$ 1.0	&...		&1.6 $\pm$ 0.6 \\
93206  &14.8 $\pm$ 1.5 	&18.3 $\pm$ 2.2 &23.2 $\pm$ 1.5 &...  		&17.7 $\pm$ 0.6	&14.1 $\pm$ 0.8	& $<$ 3.1 \\
93222  &14.8 $\pm$ 2.1 	&17.4 $\pm$ 2.4	&25.4 $\pm$ 1.9	&24.4 $\pm$ 1.6	&20.1 $\pm$ 1.3	&13.3 $\pm$ 1.8	& $<$ 6.6 \\
94493  &13.3 $\pm$ 2.0	&16.2 $\pm$ 3.6	&20.6 $\pm$ 1.6	&...  		&17.3 $\pm$ 1.1	&...		& $<$ 6.2 \\
101131 &19.7 $\pm$ 2.7	&24.5 $\pm$ 2.9 &25.7 $\pm$ 2.2	&26.6 $\pm$ 1.7	&28.3 $\pm$ 1.3	&24.7 $\pm$ 1.5	&4.2 $\pm$ 1.4 \\
101190 &22.1 $\pm$ 2.6 	&23.3 $\pm$ 2.4 &27.1 $\pm$ 2.0 &28.0 $\pm$ 2.1	&28.1 $\pm$ 1.1	&26.6 $\pm$ 1.4	&8.0 $\pm$ 1.0 \\
101413 &20.7 $\pm$ 4.1	&...		&25.4 $\pm$ 3.7 &25.2 $\pm$ 2.4	&25.7 $\pm$ 2.5	&21.1 $\pm$ 3.6	&6.8 $\pm$ 2.2 \\
101436 &18.9 $\pm$ 3.1	&20.2 $\pm$ 2.9 &20.4 $\pm$ 2.6 &21.0 $\pm$ 2.8	&20.9 $\pm$ 1.4	&18.0 $\pm$ 1.6	&10.2 $\pm$ 1.0 \\
104705 & 9.4 $\pm$ 1.5 	&12.6 $\pm$ 2.9 &13.8 $\pm$ 1.2	&14.0 $\pm$ 1.6	&13.0 $\pm$ 0.7	&...		& $<$ 3.9 \\
110432 &25.9 $\pm$ 4.0 	&27.1 $\pm$ 6.2 &30.4 $\pm$ 2.8	&30.2 $\pm$ 2.1	&34.8 $\pm$ 2.1	&32.1 $\pm$ 1.7	&13.2 $\pm$ 1.5 \\
116852 & 8.9 $\pm$ 1.2	&13.0 $\pm$ 1.4	&15.3 $\pm$ 0.9 &...  		&13.7 $\pm$ 0.7	&10.6 $\pm$ 0.8	& $<$ 6.1\\
147888 & ...		& ...		&15.6 $\pm$ 4.8	&15.1 $\pm$ 6.0	&15.6 $\pm$ 2.4	&14.2 $\pm$ 3.0	&8.8 $\pm$ 1.6 \\
148422 &23.5 $\pm$ 5.0	&21.9 $\pm$ 8.5	&22.9 $\pm$ 3.3	&26.2 $\pm$ 4.6	&26.0 $\pm$ 2.6	&...		&7.3 $\pm$ 2.1 \\
149404 &47.3 $\pm$ 7.0 	&56.7 $\pm$ 11.1&61.1 $\pm$ 7.4 &53.7 $\pm$ 3.5 &63.7 $\pm$ 3.9	&55.5 $\pm$ 4.6	&26.4 $\pm$ 2.5 \\
152233 &28.5 $\pm$ 1.9	&...		&31.1 $\pm$ 1.8	&35.4 $\pm$ 1.7	&32.0 $\pm$ 1.5	&29.1 $\pm$ 2.1	&6.5 $\pm$ 1.6 \\
152248 &27.8 $\pm$ 3.9	&...		&33.3 $\pm$ 2.2	&35.8 $\pm$ 1.8	&30.6 $\pm$ 1.8	&28.2 $\pm$ 2.2	&9.3 $\pm$ 2.1 \\
152723 &...		&41.2 $\pm$ 3.5 &46.1 $\pm$ 6.1 &46.3 $\pm$ 3.1	&49.5 $\pm$ 2.4 &43.0 $\pm$ 1.9	&6.6 $\pm$ 1.4 \\
161807 &13.5 $\pm$ 3.0	&18.3 $\pm$ 4.6	&32.1 $\pm$ 2.7	&27.7 $\pm$ 2.7	&17.4 $\pm$ 1.8	&15.3 $\pm$ 1.8	& $<$ 6.9 \\
177989 &25.5 $\pm$ 3.6 	&33.3 $\pm$ 8.8	&33.6 $\pm$ 2.6 &...  		&26.5 $\pm$ 1.9	&26.6 $\pm$ 2.7	&6.3 $\pm$ 1.6 \\
185418 &32.0 $\pm$ 6.8 	&...		&46.3 $\pm$ 6.4	&...  		&50.8 $\pm $3.9	&49.7 $\pm$ 4.2	&17.4 $\pm$ 4.9 \\
192639 &48.2 $\pm$ 7.8 	&...		&51.9 $\pm$ 6.1	&54.2 $\pm$ 4.0	&54.7 $\pm$ 3.7	&53.4 $\pm$ 4.1	&20.1 $\pm$ 4.0 \\
199579 &52.6 $\pm$ 4.3	&52.7 $\pm$ 5.5	&71.6 $\pm$ 4.5	&64.6 $\pm$ 5.3	&61.9 $\pm$ 2.6	&60.3 $\pm$ 2.7	&8.4 $\pm$ 1.2 \\
201345 &11.5 $\pm$ 0.8	&15.9 $\pm$ 1.4	&22.2 $\pm$ 1.0	&22.3 $\pm$ 1.0	&14.8 $\pm$ 0.8	&10.6 $\pm$ 0.8	& $<$ 2.4\\
206267 &42.5 $\pm$ 2.5 	&48.5 $\pm$ 9.1 &46.1 $\pm$ 5.1 &52.0 $\pm$ 3.7	&49.5 $\pm$ 2.4 &45.6 $\pm$ 3.2	&24.1 $\pm$ 2.1 \\
207198 &47.1 $\pm$ 7.8 	&53.8 $\pm$ 12.3&57.8 $\pm$ 7.1 &47.3 $\pm$ 4.6	&56.9 $\pm$ 2.8 &51.1 $\pm$ 4.3	&34.7 $\pm$ 3.2 \\
207538 &...		&...		&...		&46.6 $\pm$ 6.7	&57.4 $\pm$ 5.4 &46.6 $\pm$ 8.3	&28.0$\pm$ 3.2 \\
224151 &23.4 $\pm$ 2.7 	&...		&28.9 $\pm$ 2.0 &...  		&28.4 $\pm$ 1.5 &23.2 $\pm$ 1.5	&5.8 $\pm$ 0.7 \\
303308 &22.6 $\pm$ 2.4 	&25.3 $\pm$ 2.5 &...		&...  		&24.6 $\pm$ 1.5	&24.2 $\pm$ 2.8	&5.3 $\pm$ 1.1 \\
308813 &21.3 $\pm$ 2.5 	&26.8 $\pm$ 3.9	&38.4 $\pm$ 2.4	&35.7 $\pm$ 3.6	&31.6 $\pm$ 2.6	&24.4 $\pm$ 2.4	&4.0 $\pm$ 1.5 \\

\enddata
\end{deluxetable}

\begin{deluxetable}{llll}
\tablecolumns{4}
\tablewidth{0pc}
\tabletypesize{\scriptsize}
\tablecaption{Absorption Lines and Oscillator Strengths \label{absorption_lines}}
\tablehead{\multicolumn{2}{c} {HD Lines} &\multicolumn{2}{c} {C I*lines} \\
Wavelength (\AA)  &log(f$\lambda$)\tablenotemark{a} (cm) &Wavelength (\AA)
& log(f$\lambda$)\tablenotemark{b} (cm) }

\startdata
959.7968	&-6.8520	&945.338	&-5.842 \\
975.5524	&-6.7190	&1157.4056	&-7.759 \\
1011.4439	&-6.5769	&1157.7697	&-6.993 \\
1021.4436 	&-6.5867	&1158.5443	&-8.107 \\
1054.2800	&-6.7632	&1158.6744	&-7.770 \\
1066.2636	&-6.9114	&1158.7321	&-7.702 \\
1105.8335	&-8.0849	&1276.7498	&-7.405 \\
				&&1279.0562	&-7.676 \\
				&&1279.8907	&-6.737 \\
				&&1280.4043	&-7.249 \\
				&&1280.5975	&-7.045 \\
				&&1656.2672	&-6.004 \\
				&&1657.3792	&-6.228 \\
				&&1657.9068	&-6.103 \\
\enddata
\tablenotetext{a} {HD log(f$\lambda $) values from Abgrall \& Roueff (2006)}
\tablenotetext{b} {C I* log(f$\lambda $) values from Morton (2003) }

\end{deluxetable}

\begin{deluxetable}{lccr}
\tablecolumns{4}
\tablewidth{0pc}
\tabletypesize{\scriptsize}
\tablecaption{1105.73 \AA\ C I* f-value  \label{tab_CIfvalue}}
\tablehead{Star & $W_{1105.73}$ & log N(C I*)  &f-value\\
&(m\AA)	&(cm$^{-2}$)&(10$^{-3}$)}
\startdata
HD 12323     & 16 $\pm$ 3 & 14.52 & 6.0$^{+1.7}_{-1.4}$ \\
$o$ Per      & 17 $\pm$ 4 & 14.88 & 5.7$^{+8.0}_{-2.8}$ \\ 
HD 207538    & 32 $\pm$ 5 & 14.93 & 9.0$^{+6.3}_{-3.6}$ \\
HD 210839    & 29 $\pm$ 10 & 14.85 & 5.6$^{+3.5}_{-2.4}$ \\

Weighted Avg & & & 6.2$^{+3.5}_{-1.0}$ \\
\enddata
\end{deluxetable}

\begin{deluxetable}{llccc}
\tablecolumns{5}

\tablewidth{0pc}
\tabletypesize{\scriptsize}
\tablecaption{HD Column Density \label{colden}}
\tablehead{Star (HD)  & \multicolumn{3}{c} {Single Component COG}
&\multicolumn{1}{c} {Multiple Component COG} \\ &log N(HD) &$b$-value &
&log N(HD)  }

\startdata
24534	&15.88 $\pm$ 0.40\tablenotemark{a}	&1.8 $\pm$ 0.2	&	&15.71 $\pm$ 0.07 \\
27778	&15.89 $\pm$ 1.10			&2.2 $\pm$ 0.6	&	&15.84 $\pm$ 0.12 \\
53367	&15.74 $\pm$ 0.34			&4.0 $\pm$ 0.7	&	&16.23 $\pm$ 0.17 \\
73882	&15.81 $\pm$ 1.18			&2.9 $\pm$ 1.1	&	&16.03 $\pm$ 0.34 \\
110432	&15.43 $\pm$ 0.12\tablenotemark{a}	&2.3 $\pm$ 0.2	&	&15.30 $\pm$ 0.08 \\
147888	&15.39 $\pm$ 1.47			&1.0 $\pm$ 0.4	&	&15.30 $\pm$ 0.15 \\
149404	&15.74 $\pm$ 0.12\tablenotemark{a}	&4.1 $\pm$ 0.3	&	&15.50 $\pm$ 0.18 \\
185418	&15.51 $\pm$ 0.32\tablenotemark{a}	&3.7 $\pm$ 0.6	&	&15.87 $\pm$ 0.32 \\
192639	&15.57 $\pm$ 0.22\tablenotemark{a}\tablenotemark{b}	&4.0 $\pm$ 0.4	&	&15.10 $\pm$ 0.14 \\
199579	&15.08 $\pm$ 0.09\tablenotemark{a}	&6.5 $\pm$ 0.7	&	&15.09 $\pm$ 0.02 \\
206267	&15.72 $\pm$ 0.14\tablenotemark{a}	&3.4 $\pm$ 0.2	&	&15.54 $\pm$ 0.10 \\
207198	&16.12 $\pm$ 0.47\tablenotemark{a}	&3.4 $\pm$ 0.4	&	&15.84 $\pm$ 0.08 \\
207538	&15.83 $\pm$ 0.25\tablenotemark{a}	&3.6 $\pm$ 0.5	&	&15.82 $\pm$ 0.09 \\	
 \hline \\ \hline

12323	&15.02 $\pm$ 0.18			&5.4 $\pm$ 0.8	& &\\
15558	&15.64 $\pm$ 0.32			&6.4 $\pm$ 1.4	& &\\
45314	&15.30 $\pm$ 0.37\tablenotemark{a}	&4.9 $\pm$ 0.7	& &\\
74920	&15.33 $\pm$ 0.04\tablenotemark{a}	&3.2 $\pm$ 0.1	& &\\
90087	&14.30 $\pm$ 0.07			&5.7 $\pm$ 2.6	& &\\
91651	&13.43 $\pm$ 0.45			&$\ge$0.6	& &\\
91824	&14.17 $\pm$ 0.07\tablenotemark{a}	&$\ge$4.8 	& &\\
93204	&14.22 $\pm$ 0.10\tablenotemark{a}	&$\ge$4.2 	& &\\
93205	&14.28 $\pm$ 0.07			&$\ge$5.7	& &\\
93206	&14.19 $\pm$ 0.10			&4.1 $\pm$ 2.7 	& &\\
93222	&14.19 $\pm$ 0.12			&$\ge$3.2	& &\\
94493	&14.24 $\pm$ 0.32			&$\ge$1.3	& &\\
101131	&14.85 $\pm$ 0.19\tablenotemark{a}	&2.4 $\pm$ 0.3	& &\\
101190	&15.10 $\pm$ 0.12			&2.2 $\pm$ 0.1	& &\\
101413	&15.01 $\pm$ 0.31			&2.0 $\pm$ 0.3	& &\\
101436	&15.36 $\pm$ 0.17			&1.4 $\pm$ 0.5	& &\\
104705	&14.26 $\pm$ 0.53			&1.5 $\pm$ 1.2	& &\\
116852	&14.15 $\pm$ 0.13			&2.0 $\pm$ 1.5	& &\\
148422	&15.07 $\pm$ 0.28			&2.0 $\pm$ 0.3	& &\\
152233	&14.95 $\pm$ 0.19			&2.9 $\pm$ 0.3	& &\\
152248	&15.15 $\pm$ 0.18\tablenotemark{a}	&2.6 $\pm$ 0.3	& &\\
152723	&15.01 $\pm$ 0.13\tablenotemark{a}	&4.5 $\pm$ 0.5	& &\\
161807	&14.12 $\pm$ 0.07\tablenotemark{a}	&$\ge$4.5 	& &\\
177989	&14.92 $\pm$ 0.22\tablenotemark{a}	&2.6 $\pm$ 0.4	& &\\
201345	&14.00 $\pm$ 0.04			&$\ge$5.8	& &\\
224151	&14.91 $\pm$ 0.09\tablenotemark{a}	&2.4 $\pm$ 0.2	& &\\
303308	&14.90 $\pm$ 0.14			&2.2 $\pm$ 0.3	& &\\
308813	&14.48 $\pm$ 0.14			&5.2 $\pm$ 2.1	& &\\

\enddata
\tablenotetext{a}{Assumed a good fit then scaled errors to a reduced chi
square of one.}\tablenotetext{b}{The single COG is a better solution and
agrees with the Profile fitting column densities within errors while 
 the multiple component COG is not within errors and misfit the data (see Figure
\ref{plotthree}).}  

\end{deluxetable}

\begin{deluxetable}{lcccc}
\tablecolumns{5}
\tablewidth{0pc}
\tabletypesize{\scriptsize}
\tablecaption{Column Density Comparison \label{profile_vsCOG}}
\tablehead{Star &\multicolumn{2}{c} {Profile Fitting} &\multicolumn{2}{c}{Curve of Growth}\\
	&log N(HD)	&$b$-value	&log N(HD)	&$b$-value\\
& (cm$^{-2}$)&&(cm$^{-2}$)& }

\startdata
90087	&14.36 $\pm$ 0.05  &3.59	&14.30 $\pm$ 0.07	&5.7 \\
101436 	&15.12 $\pm$ 0.08  &1.33	&15.36 $\pm$ 0.17	&1.4 \\
177989 	&14.92 $\pm$ 0.07  &2.59	&14.92 $\pm$ 0.22	&2.6 \\
110432  & 15.32 $\pm$ 0.04 &2.32	&15.43 $\pm$ 0.12	&2.3\\
192639  & 15.52 $\pm$ 0.07 &3.70	&15.57 $\pm$ 0.22	&4.0 \\
\enddata
\end{deluxetable}

\begin{deluxetable}{lcccc}
\tablecolumns{5}
\tablewidth{0pc}
\tabletypesize{\scriptsize}
\tablecaption{Correlation Coefficients \label{correlate}}
\tablehead{Quantities	&$r$\tablenotemark{a}& $\rho$ & $\#$ &Slope}
\startdata
log N(HD) vs. log N(H$_2$)	&0.94	&0.89 : 0.96	&55	&1.25 $\pm$ 0.03 \\
log N(HD) vs. log N(CO)		&0.93	&0.77 : 0.98	&13	&0.49 $\pm$ 0.02 \\
log N(HD) vs. log N(H$_{Tot}$)	&0.67	&0.48 : 0.80	&48	&1.80 $\pm$ 0.05 \\
log N(HD) vs. E(B-V)		&0.78	&0.64 : 0.87	&51	&3.94 $\pm$ 0.09 \\
log N(H$_2$) vs. E(B-V)		&0.75	&0.59 : 0.85	&51	&2.34 $\pm$ 0.04 \\
log N(HD) vs. A$_V$		&0.56	&0.27 : 0.75	&35	&...\\
log N(H$_2$) vs. A$_V$		&0.53	&0.23 : 0.73	&35	&...\\

log N(Fe II/H$_{Tot}$) vs. log N(HD)		&-0.78	&-0.92 : -0.44 &15	&-0.30 $\pm$ 0.07 \\
log N(Fe II/H$_{Tot}$) vs. log N(HD/H$_{Tot}$)  &-0.72	&-0.90 : -0.33 &15	&-0.33 $\pm$ 0.09 \\
log N(Fe II/H$_{Tot}$) vs. log N(HD/2H$_2$)	&-0.80	&-0.93 : -0.48 &15	&-0.85 $\pm$ 0.17 \\
\hline $\qquad$Correlations not shown\\ \hline
log N(HD) vs. log N(H I)		&0.34	&0.07 : 0.57	&48	&...	\\
log N(H$_2$) vs. log N(H I)		&0.33	&0.05 : 0.56	&48	&...	\\
log N(HD) vs. Rv			&-0.25\tablenotemark{b}	& ... :	-0.54	&35	&...	\\
log N(H$_2$) vs. Rv			&-0.26\tablenotemark{b}	& ... : -0.54	&35	&...	\\
$f_{H2}$ vs. log N(HD)		&0.74	&0.58 : 0.85	&48	&... \\
$f_{H2}$ vs. log N(H$_2$) 	&0.81	&0.68 : 0.89	&48	&... \\

\enddata
\tablenotetext{a}{r is  the Pearson correlation coefficient, $\rho$ is the
95$\%$ confidence interval(uncertainties in the correlation coefficients),
$\#$ is the number of data points used in the correlation, and 
Slope is the slope of the weighted least-squares line fit.}
\tablenotetext{b}{Correlations not 95$\%$ significant}

\end{deluxetable}

\clearpage

\clearpage

\clearpage

\clearpage

\clearpage

\clearpage

\end{document}